\documentclass[aps,twocolumn,preprintnumbers,nofootinbib,superscriptaddress]{revtex4}
\usepackage{amsmath} \usepackage{graphicx} \usepackage{amsfonts}
\usepackage{array} \usepackage{amsthm} \usepackage{bm}
\usepackage{palatino} \usepackage{mathpazo} %\usepackage{mathptmx}
\usepackage{supertabular}
\pdfoutput=1

\usepackage[breaklinks]{hyperref}
\usepackage{color}

\definecolor{oxfordblue}{rgb}{0.0, 0.13, 0.28}
\definecolor{burgundy}{rgb}{0.5, 0.0, 0.13}
\definecolor{darkolivegreen}{rgb}{0.33, 0.42, 0.18}
\definecolor{darkblue}{rgb}{0,0,0.5}
\definecolor{richcarmine}{rgb}{0.84, 0.0, 0.25}
\definecolor{bluer}{rgb}{0.00,0.50,0.75}{}
\hypersetup{colorlinks=true, citecolor=red, linkcolor=blue,
urlcolor = magenta, filecolor=magenta}

\newcommand{\be}{\begin{equation}}
\newcommand{\ee}{\end{equation}}
\newcommand{\ba}{\begin{eqnarray}}
\newcommand{\ea}{\end{eqnarray}}
\newcommand{\bal}{\begin{align}}
\newcommand{\eal}{\end{align}}

\newcommand{\dd}{\text{d}}

\newcommand{\bw}{\begin{widetext}}
\newcommand{\ew}{\end{widetext}}

\begin{document}

\title{Chaotic imprints of dark matter in extreme mass-ratio inspirals}

\author{Mustapha Azreg-A\"{i}nou}\email{azreg@baskent.edu.tr}
\affiliation{Ba\c{s}kent University, Engineering Faculty, Ba\u{g}lica Campus,
06790-Ankara, T\"{u}rkiye}

\author{Mubasher Jamil}\email{mjamil@sns.nust.edu.pk}
\affiliation{School of Natural Sciences, National University of Sciences and 
Technology, H-12, Islamabad 44000, Pakistan}
%\affiliation{Research Center of Astrophysics and Cosmology, Khazar University, 
%Baku, AZ 1096, 41 Mehseti Street, Azerbaijan}

\author{Emmanuel N. Saridakis} \email{msaridak@noa.gr}
\affiliation{Institute for Astronomy, Astrophysics, Space Applications and 
Remote Sensing, National Observatory of Athens, 15236 Penteli, Greece}
\affiliation{Departamento de Matem\'{a}ticas, Universidad Cat\'{o}lica del 
Norte, Avda. Angamos 0610, Casilla 1280 Antofagasta, Chile}
\affiliation{CAS Key Laboratory for Researches in Galaxies and Cosmology, 
School 
of Astronomy and Space Science, University of Science and Technology of China, 
Hefei, Anhui 230026, China}

\begin{abstract}
Extreme mass-ratio inspirals (EMRIs) are among the most powerful probes of 
strong-field gravity and of the environments surrounding supermassive compact 
objects. Motivated by the expected presence of dark matter near galactic 
centers, we investigate the emergence and gravitational-wave imprints of 
chaotic dynamics in EMRIs evolving in non-vacuum spacetimes. Within a unified dynamical framework, we analyze test-particle motion in a broad class ofdark-matter-embedded geometries, including singular black holes, 
regular black holes, naked singularities, and Einstein-cluster configurations. We show that environmental perturbations generically break integrability in the strong-field regime, giving rise to chaotic motion whose onset, duration, and termination depend sensitively on horizon structure, core regularization, and matter distribution. Using the numerical Kludge approach, we demonstrate that chaotic trajectories produce systematic qualitative modifications of the emitted gravitational radiation, such as irregular amplitude modulation and loss of phase coherence, in contrast to the smooth, quasi-periodic waveforms generated by regular motion. Our results establish the robustness of chaos in environmentally perturbed EMRIs and provide a clear conceptual link between nonlinear orbital dynamics, spacetime structure, and observable gravitational-wave signatures.
\end{abstract}

%\pacs{}

\maketitle

\section{Introduction}

The direct detection of gravitational waves (GWs) by the LIGO and Virgo
collaborations marked a turning point in modern astrophysics, placing
gravitational-wave astronomy as a precision observational discipline
\cite{Abbott:2016blz,Abbott:2019yzh}. Gravitational waves provide a
fundamentally new probe of the strong-field regime of gravity, granting
direct access to dynamical spacetime phenomena that are either invisible or
only indirectly inferred through electromagnetic observations.
Among the most promising future sources in this context are
\emph{extreme mass-ratio inspirals} (EMRIs), consisting of a stellar-mass
compact object spiraling into a supermassive compact object residing at the
center of a galaxy.

EMRIs are expected to be primary targets of forthcoming space-based
gravitational-wave detectors such as LISA \cite{LISA:2017pwj},
TianQin \cite{Luo:2015ght}, and Taiji \cite{Ruan:2018tsw}, operating in the
millihertz frequency band. A defining feature of EMRIs is the exceptional
duration of their signals, which can span $\mathcal{O}(10^4$-$10^6)$ orbital
cycles before plunge. This extreme phase coherence renders EMRIs uniquely
sensitive probes of spacetime geometry, enabling high-precision tests of
general relativity \cite{Barack:2018yly}, spacetime multipolar structure
\cite{Ryan:1995wh,Glampedakis:2005hs}, alternative theories of gravity
\cite{Yunes:2011we,Berti:2015itd,CANTATA:2021asi}, and environmental effects 
surrounding
supermassive compact objects \cite{Kocsis:2011dr,Barausse:2014tra}.

In realistic astrophysical settings, EMRIs do not evolve in vacuum.
Supermassive black holes are expected to be embedded in complex environments
that may include accretion disks \cite{Narayan:2005ie}, stellar clusters
\cite{Merritt:2013}, magnetic fields \cite{Blandford:2018iot}, and
dark-matter distributions \cite{Bertone:2004pz,Gnedin:2003rj}.
Although such effects are often treated as small perturbations, their
cumulative impact over the long inspiral timescales characteristic of EMRIs
can become dynamically significant. In particular, even weak environmental
forces may break the integrability of geodesic motion, potentially leading
to chaotic dynamics.

Chaos, characterized by sensitive dependence on initial conditions and the
absence of global constants of motion, is a generic feature of nonlinear
Hamiltonian systems. In relativistic gravity, chaotic behavior has been
identified in a wide range of contexts, including test-particle motion in
deformed or perturbed black-hole spacetimes
\cite{Cornish:1996hx,Levin:1999zx,Contopoulos:2002}, compact-object binaries
\cite{Schnittman:2007ij}, and relativistic systems subject to external fields
or non-vacuum matter configurations \cite{Semerak:2010lzj}.
In the EMRI context, chaos is of particular interest because it can disrupt
the quasi-periodic structure of the orbit and, in turn, leave distinctive
imprints on the emitted gravitational radiation.

Motivated by astrophysical evidence for dark-matter concentrations near
galactic centers \cite{Ghez:2008ms,Schoedel:2009mv}, numerous new exact 
solutions of black hole surrounded by dark matter have been reported in the 
literature 
\cite{Cruz:2017ecg,Rizwan:2025pvp,Rayimbaev:2021kjs,Nampalliwar:2021tyz,
Jusufi:2020cpn, 
Hendi:2020zyw,Rizwan:2018rgs,Mach:2021zqe,Baggioli:2021ejg,Gonzalez:2023rsd,
Al-Badawi:2025njy, 
Yang:2024lmj,Barrientos:2024umq, 
Santos:2025fdp,Anand:2025rjg,Anand:2025cer,Nozari:2026wjo}.  
Recent studies have
explored the emergence of chaotic dynamics in EMRIs embedded in dark-matter
environments. In particular, Refs.~\cite{Das:2025vja,Das:2025eiv} examined the
motion of a stellar-mass object subject to an effective harmonic confinement
and orbiting a Schwarzschild-like black hole surrounded by a Dehnen-type
dark-matter halo. Using orbital projections, Poincar\'e sections, and
Lyapunov indicators, these works demonstrated the existence of transitions
from regular to chaotic motion controlled by the orbital energy and
dark-matter parameters. Furthermore, qualitative differences between
gravitational-wave signals generated by chaotic and non-chaotic trajectories
were identified, including amplitude irregularities and spectral
broadening, as reported in Ref. \cite{Cornish:2003uq}. It is interesting to 
note that both periodic 
as well as chaotic motions of neutral and charged particles around black holes 
have been frequently investigated in the literature 
\cite{Bombelli:1991eg,Sota:1995ms,Vieira:1996zf,Suzuki:1996gm,Han:2008zzf,Chen:2016tmr,Wang:2022kvg,Tu:2023xab,Shabbir:2025kqh}.

Despite this progress, several key questions remain open.
Existing analyses have focused almost exclusively on singular,
Schwarzschild-like spacetimes and on specific realizations of dark-matter
halos, leaving unexplored the extent to which chaotic dynamics and their
gravitational-wave signatures are robust under changes in the underlying
spacetime geometry.
In particular, it remains unclear which aspects of chaos constitute
universal  features of environmentally perturbed EMRIs, and which depend
sensitively on the presence of horizons, curvature singularities, or
regularized cores.
This issue is especially timely in view of the growing interest in
\emph{regular black holes}, which resolve central singularities while
preserving asymptotic black-hole properties and arise in a variety of
theoretical frameworks
\cite{Bardeen:1968,Hayward:2005gi,Bambi:2013ufa,Fan:2016hvf,
Abdujabbarov:2016hnw,Borde:1996df,Carballo-Rubio:2018pmi,Lan:2023cvz,
Bronnikov:2006fu,Simpson:2019mud,Lemos:2011dq,Berej:2006cc,Kumar:2019pjp,
Flachi:2012nv,Li:2013jra,Ovgun:2019wej,Burinskii:2002pz,Neves:2014aba,
Jusufi:2018jof,Toshmatov:2017zpr,Bueno:2024dgm,
Dymnikova:2004zc,Bronnikov:2005gm,Jusufi:2022rbt,
Simpson:2018tsi}.
Since the near-horizon and interior structure of spacetime can directly
influence orbital stability, capture mechanisms, and the duration of chaotic
phases, a comparative analysis across different compact-object geometries is
essential for assessing the physical relevance of chaos in realistic EMRI
systems.

The purpose of the present work is to address these issues through a
systematic and comparative investigation of chaotic dynamics in EMRIs
embedded in dark-matter spacetimes.
Adopting a unified dynamical framework based on horizon-regular coordinates,
we analyze test-particle motion in a broad class of spherically symmetric
geometries, including singular black holes, regular black holes, naked
singularities, and Einstein-cluster configurations.
Environmental effects are modeled through controlled external perturbations
that explicitly break integrability, allowing us to isolate the onset,
duration, and termination of chaotic motion across different spacetime
realizations.

In addition to characterizing the orbital dynamics, we examine the
gravitational-wave signatures of chaos using the numerical Kludge approach.
While approximate in nature, this method is well suited for identifying
robust qualitative imprints of chaotic motion on the emitted radiation.
Our analysis demonstrates that chaos generically leads to irregular amplitude
modulation and loss of waveform coherence, and that these features persist
across a wide range of spacetime geometries, albeit with important
differences linked to horizon structure and core regularization.

The structure of the paper is as follows.
In Section \ref{sec:framework} we introduce the dynamical framework and 
describe the class of
spherically symmetric spacetimes under consideration.
Section \ref{sec:chaos} presents the diagnostics employed to identify chaotic 
behavior.
In Section \ref{sec:metrics} we perform a comparative analysis of chaotic 
dynamics across
singular, regular, and horizonless dark-matter-embedded spacetimes.
Section \ref{Gwswction} is devoted to the corresponding gravitational-wave 
signatures and
their qualitative imprints.
Finally, Section  \ref{secconc}   summarizes our results and outlines 
directions for future
work.

\section{Dynamical Framework for EMRIs in Dark Matter Spacetimes}
\label{sec:framework}

In this section we establish the dynamical framework employed throughout the
paper to describe the motion of a compact object in the vicinity of a massive
central configuration embedded in a dark-matter environment. Our aim is to
formulate the equations of motion in a manner that is sufficiently general to
accommodate a wide class of spherically symmetric spacetimes,  while remaining 
suitable for long-term numerical evolution in
the strong-field regime.

A central requirement for the present analysis is the ability to follow the
particle dynamics across regions where standard Schwarzschild-like coordinates
become ill-behaved, such as near horizons or effective capture surfaces. For
this reason, we adopt horizon-regular coordinate systems that allow for a
continuous description of the motion and a transparent Hamiltonian formulation.
This choice ensures that the onset of chaotic behavior, as well as its finite
duration in spacetimes admitting capture, can be analyzed without coordinate
artifacts.

Environmental effects associated with the surrounding dark-matter distribution
are incorporated through effective external forces acting on the orbiting
object. While simplified in form, this approach provides a controlled mechanism
for breaking integrability and isolating the impact of non-vacuum environments
on the orbital dynamics. The resulting framework serves as the basis for the
chaos diagnostics discussed in the following section and for the construction
of gravitational-wave signals analyzed later in the paper.

\subsection{Spherically symmetric spacetimes and horizon-regular coordinates}
\label{subsec:sss}

We begin by considering a generic spherically symmetric and static (SSS) 
spacetime described by the line element
\begin{equation}
\label{eq:sss}
\dd s^2 = -c^2 f(r)\,\dd t^2 + \frac{\dd r^2}{f(r)} + r^2 \dd\Omega^2 \, ,
\end{equation}
where $\dd\Omega^2=\dd\theta^2+\sin^2\theta\,\dd\phi^2$  is the metric on the 
unit two-sphere.
Such geometries encompass a wide class of black-hole and compact-object 
solutions, including Schwarzschild, regular black holes, and effective 
spacetimes describing dark-matter environments.

For the numerical investigation of particle dynamics, Schwarzschild-like 
coordinates are not suitable when the metric function $f(r)$ vanishes at a 
horizon. In order to regularize the metric at such locations, we introduce 
horizon-regular Painlev\'{e}-Gullstrand-type coordinates. For the metric 
\eqref{eq:sss}, the transformation
\begin{align}
c\,\dd t &= c\,\dd t_{\rm new} - \frac{\sqrt{1-f(r)}}{f(r)}\,\dd r_{\rm new} 
\,, 
\nonumber\\
\dd r &= \dd r_{\rm new} \, ,
\end{align}
brings the line element into the form
\begin{equation}
\label{eq:ssspg}
\dd s^2 = -c^2 f(r)\,\dd t^2 + 2c\sqrt{1-f(r)}\,\dd t\,\dd r + \dd r^2 + 
r^2\dd\Omega^2 \, ,
\end{equation}
where we have omitted the label {\textgravedbl}new{\textacutedbl} for 
simplicity.

More generally, we consider the SSS metric
\begin{equation}
\label{eq:sssg}
\dd s^2 = -c^2 f(r)\,\dd t^2 + \frac{\dd r^2}{g(r)} + r^2 \dd\Omega^2 \, .
\end{equation}
In this case, a generalized Painlev\'{e}-Gullstrand coordinate transformation
\begin{align}
\label{eq:pgtrans}
c\,\dd t &= c\,\dd t_{\rm new}
- \frac{\sqrt{1-g(r)}}{\sqrt{f(r)g(r)}}\,\dd r_{\rm new} \,, 
\nonumber\\
\dd r &= \dd r_{\rm new} \, ,
\end{align}
yields the horizon-regular metric
\begin{equation}
\label{eq:sssgpg}
\dd s^2 = -c^2 f(r)\,\dd t^2 + 2c\,h(r)\sqrt{1\!-\!g(r)}\,\dd t\,\dd r + \dd 
r^2 + 
r^2\dd\Omega^2   ,
\end{equation}
where $h(r)\equiv \sqrt{f(r)/g(r)}$. Since the functions $f$ and $g$ have the 
same roots, the function $h$ is horizon-regular. Here again we have dropped the 
subscript {\textgravedbl}new{\textacutedbl}.

\subsection{Effective Lagrangian and external environmental forces}
\label{subsec:lagrangian}

We now consider the motion of a test particle of mass $m$ confined to the 
equatorial plane $\theta=\pi/2$ of the spacetime \eqref{eq:sssgpg}. In the 
presence of environmental effects, modeled effectively through external forces, 
the dynamics is described by the Lagrangian
\begin{multline}
\label{eq:lagrangian}
\mathcal{L}
= -mc\sqrt{c^2 f(r) - 2c\,h(r)\sqrt{1-g(r)}\,\dot r - \dot r^{\,2} - r^2 
\dot\phi^{\,2}} \\
- \frac{k_r}{2}\,(r-r_c)^2 - \frac{k_\phi r_h^2}{2}\,\phi^2 \, ,
\end{multline}
where overdots denote derivatives with respect to coordinate time $t$, and 
$k_r$, $k_\phi$, and $r_c$ are positive constants.

The last two terms in \eqref{eq:lagrangian} represent effective harmonic 
potentials in the radial and angular directions. Such terms provide a simple 
phenomenological description of environmental perturbations, such as the 
effect of surrounding dark matter or stellar backgrounds, and have been 
widely employed in the literature as a controlled way to introduce 
non-integrability in otherwise symmetric systems. 
In the present analysis these terms should be understood as 
local effective perturbations rather than as a microscopic model of a specific 
dark-matter interaction. They are meant to mimic, in a controlled 
phenomenological way, environmental effects that can perturb the motion in the 
strong-field region, such as dynamical friction, local inhomogeneities, stellar 
perturbations, or deviations from exact spherical symmetry. Dynamical friction, 
for instance, plays an important role in the interaction between dark matter 
and 
moving astrophysical objects~\cite{Chandra43,Dosopoulou24}, and may affect 
orbital quantities such as the semi-major axis and eccentricity. Recent studies 
have also shown that self-interacting dark matter can enhance such effects 
\cite{Fischer24,Glennon24}. In a realistic system these environmental 
perturbations would generally depend on the local density, velocity 
distribution, and orbital parameters, and would therefore decay away from the 
region where the environment is dynamically relevant. Accordingly, the harmonic 
form adopted in \eqref{eq:lagrangian} should be regarded as a local 
approximation valid near the characteristic radius $r_c$, rather than as a 
global force profile extending to arbitrarily large radii. 
In the present work, these potentials are used solely as an effective modeling 
tool, and no claim is made regarding a detailed microscopic description of the 
underlying environment.

The validity of this effective description is restricted to regimes where the 
external forces act as small perturbations on the geodesic motion, allowing one 
to isolate and study qualitative features of the resulting dynamics.

\subsection{Equations of motion and conserved quantities}
\label{subsec:eom}

We now proceed to derive the equations of motion associated with the
Lagrangian~\eqref{eq:lagrangian}. The corresponding canonical momenta and
Hamiltonian structure provide a convenient formulation of the dynamics,
allowing for a systematic investigation of conserved quantities and
integrability properties. The canonical momenta conjugate to 
$r$ and $\phi$ are obtained as
\begin{equation}
p_r = \frac{\partial \mathcal{L}}{\partial \dot r} = \frac{mc[\dot r 
+c\,h(r)\sqrt{1-g(r)}\,]}{\sqrt{c^2 f(r) - 2c\,h(r)\sqrt{1-g(r)}\,\dot r - \dot 
	r^{\,2} - r^2 
	\dot\phi^{\,2}}} \, ,
\end{equation}
\begin{equation}
p_\phi = \frac{\partial \mathcal{L}}{\partial \dot \phi} = 
\frac{mcr^2\dot\phi}{\sqrt{c^2 f(r) - 2c\,h(r)\sqrt{1-g(r)}\,\dot r - \dot 
	r^{\,2} - r^2 \dot\phi^{\,2}}} \, .
\end{equation}
The conserved energy of the test mass $m$ is given by
\begin{multline}
E = p_r \dot r + p_\phi \dot\phi - \mathcal{L}\\ = 
\frac{mc^2[cf(r)-h(r)\sqrt{1-g(r)}\,\dot r]}{\sqrt{c^2 f(r) - 
	2c\,h(r)\sqrt{1-g(r)}\,\dot r - \dot r^{\,2} - r^2 
	\dot\phi^{\,2}}} \\ + \frac{k_r}{2}\,(r-r_c)^2 + 
\frac{k_\phi r_h^2}{2}\,\phi^2 \,, 
\end{multline}
which after eliminating ($\dot r,\,\dot\phi$) in terms of ($p_r,\,p_\phi$), 
reads
\begin{multline}\label{energy}
E=-c\,h(r)\sqrt{1-g(r)}\,p_r + \frac{c\,h(r)\sqrt{(c^2m^2+p_r^2)r^2 + 
	p_\phi^2}}{r}\\ + 
\frac{k_r}{2}\,(r-r_c)^2 + \frac{k_\phi r_h^2}{2}\,\phi^2 \,.
\end{multline}

The equations of motion can be written in first-order Hamiltonian form as
\begin{equation}
\dot r = -c\,h(r)\sqrt{1-g(r)} + \frac{c\,h(r)rp_r}{\sqrt{(c^2m^2+p_r^2)r^2 + 
	p_\phi^2}} \, 
,
\end{equation}
\begin{equation}
\dot \phi = \frac{c\,h(r)p_\phi}{r\sqrt{(c^2m^2+p_r^2)r^2 + p_\phi^2}} \, ,
\end{equation}
\begin{multline}\label{eq:pr}
\dot p_r =c\,h(r) \Big[-\frac{g'(r)p_r}{2\sqrt{1-g(r)}} + 
\frac{p_\phi^2}{r^2\sqrt{(c^2m^2+p_r^2)r^2 + p_\phi^2}}\Big]\\
+\frac{c\,h'(r)[rp_r\sqrt{1-g(r)} - \sqrt{(c^2m^2+p_r^2)r^2 + p_\phi^2}]}{r}\\ 
-k_r (r-r_c) \, ,
\end{multline}
\begin{equation}
\dot p_\phi = -k_\phi r_h^2 \phi \, .
\end{equation}
It is worth mentioning that in the case where $g(r)=f(r)$, that is 
$h(r)=\sqrt{f(r)/g(r)}\equiv 1$, the second line in~\eqref{eq:pr} vanishes 
identically and all equations of motion reduce to those of the case 
$g(r)=f(r)$.

In the numerical analysis presented below, we adopt units in which $M=1=c=G$, 
while the mass of the  secondary compact object  is 
fixed to 
$m=10^{-5}$. The outer horizon radius is denoted by $r_h$, and for the external 
potentials we choose $k_r=100$, $k_\phi=25$, and $r_c=4$. In cases where the 
central object corresponds to a naked singularity or a regular compact object 
without a horizon, $r_h$ in \eqref{eq:lagrangian} is replaced by the 
corresponding characteristic radius of the object. Since the 
external perturbations introduced in
\eqref{eq:lagrangian} are treated phenomenologically, the parameters
$(k_r,\,k_\phi)$ are not fixed by a specific microscopic dark-matter model and
are therefore not directly correlated with the parameters characterizing the
background spacetime geometries considered in the following sections. In the
present work, the adopted values of $(k_r,\,k_\phi)$ are chosen so as to
provide controlled non-integrable perturbations capable of generating chaotic
dynamics, thereby allowing a comparative investigation of chaos across
different compact-object configurations.

Due to the presence of the additional potential terms in the
Lagrangian~\eqref{eq:lagrangian}, the resulting trajectories are, in general,
non-geodesic.

\section{Diagnostics of Chaotic Dynamics}
\label{sec:chaos}

Chaotic dynamics in relativistic systems is characterized by sensitive 
dependence on initial conditions and by the breakdown of long-term 
predictability, even when the underlying equations of motion are fully 
deterministic. In the context of extreme mass-ratio inspirals embedded in 
non-vacuum environments such as dark matter halos, chaos may arise from the 
interplay 
between strong-field gravity, external perturbations, and environmental 
structure.

In the spacetimes considered in the present work, 
the
additional perturbative potentials introduced in the Lagrangian generally break
the integrability of the underlying orbital dynamics, allowing the emergence of
chaotic motion in the strong-field regime. In order to characterize this
behavior, we employ complementary diagnostics sensitive to different aspects of
non-integrable dynamics. In particular, Lyapunov exponents provide a
quantitative measure of the sensitivity of nearby trajectories to initial
conditions, while Poincar\'e sections offer a geometric visualization of the
structure of phase space and the transition from regular to chaotic motion.
Irregularly distributed points and the destruction of invariant curves in the
reduced phase space constitute characteristic signatures of chaos, as will be
demonstrated in the following analysis.  Since no single diagnostic provides a 
complete characterization of 
chaotic behavior, it is essential to employ a combination of complementary 
tools that probe both the geometrical and dynamical aspects of the motion.

In this section, we introduce and apply a set of standard diagnostics of chaos 
adapted to the present framework. These include projections of orbital motion 
and Poincar\'e sections, which offer direct insight into phase-space structure, 
as well as criteria for identifying transitions between regular and chaotic 
regimes. We place particular emphasis     on the finite-time nature of 
chaotic 
behavior in strong-field spacetimes, where capture by a horizon or singularity 
generically limits the duration of chaotic motion. The diagnostics presented 
here form the basis for the comparative analysis of different spacetime 
geometries in Section \ref{sec:metrics} and for the investigation of 
gravitational-wave 
signatures discussed in Section \ref{Gwswction}.

\subsection{Orbital projections and Poincar\'e sections}
\label{subsec:poincare}

A first qualitative characterization of the orbital dynamics is obtained by 
examining projections of particle trajectories onto the equatorial plane 
$\theta=\pi/2$. In integrable or quasi-integrable regimes, such projections 
exhibit smooth, regular patterns corresponding to bounded and recurrent motion. 
In contrast, chaotic trajectories typically generate irregular, densely 
populated regions in configuration space, reflecting the sensitive dependence 
on 
initial conditions and the absence of additional constants of motion beyond 
those imposed by symmetries.

While orbital projections provide an intuitive visualization of the motion, a 
more refined diagnostic is furnished by Poincar\'e sections. These are 
constructed by recording the phase-space coordinates of the particle each time 
the trajectory intersects a prescribed surface of section. In the present 
analysis, we employ sections in the $(r,p_r)$ and $(\phi,p_\phi)$ planes, 
evaluated at fixed values of the conjugate angular or radial variables, 
respectively.

For regular motion, Poincar\'e sections consist of closed invariant curves, 
indicating that the dynamics is confined to tori in phase space. Chaotic 
motion, 
on the other hand, is characterized by the progressive destruction of such 
invariant structures and the appearance of scattered points that fill extended 
regions of the section. Intermediate configurations, where invariant curves 
coexist with stochastic layers, signal the onset of chaos and the breakdown of 
integrability.

\subsection{Transition to chaos and finite-time chaotic behavior}
\label{subsec:transition}

In the systems under consideration, the transition from regular to chaotic 
dynamics does not necessarily occur abruptly. Instead, particle motion often 
displays a mixed character, whereby initially regular trajectories evolve into 
chaotic ones after a finite time interval. Such behavior is commonly 
encountered 
in non-integrable Hamiltonian systems and is particularly relevant in 
relativistic settings where strong-field effects become significant.

A salient feature of the present framework is the emergence of finite-time 
chaos. In this case, the motion exhibits clear chaotic signatures, such as 
stochastic Poincar\'e sections and irregular orbital projections, over a 
limited 
time span, before the trajectory terminates due to capture by the central 
object. This phenomenon is especially pronounced in regions close to horizons 
or 
singularities, where the effective potential and spacetime curvature vary 
rapidly and amplify nonlinear effects.

Near-horizon dynamics plays a crucial role in this context. As the particle 
approaches the strong-field region, small perturbations in the initial 
conditions can lead to markedly different orbital evolutions, enhancing 
sensitivity and promoting chaotic behavior. Nevertheless, since the inspiraling 
object ultimately crosses the horizon or reaches a singularity, chaos manifests 
itself only over a finite duration. For this reason, the relevant notion is not 
asymptotic chaos in the mathematical sense, but rather finite-time chaotic 
dynamics with direct physical implications for the orbital evolution.

In this work, we adopt a practical operational definition of chaos based on the 
combined use of orbital projections and Poincar\'e sections. Trajectories are 
classified as chaotic when their projections lose regular structure and the 
corresponding Poincar\'e sections exhibit scattered, non-toroidal distributions 
over the available phase-space region. This criterion allows for a consistent 
identification of chaotic behavior across different spacetime geometries and 
parameter choices, and it is particularly well suited for systems where the 
motion terminates after a finite time.

\section{Dark-Matter-Embedded Spacetimes: A Comparative Analysis of Chaotic 
Dynamics}
\label{sec:metrics}

In this section we present the core comparative analysis of chaotic dynamics
in extreme mass-ratio inspirals evolving within dark-matter-embedded
spacetimes. Our central aim is to determine which features of chaotic motion
are generic consequences of environmental perturbations in the strong-field
regime, and which depend sensitively on the geometric structure of the
underlying spacetime.

To address this question systematically, we examine representative geometries
that span a wide spectrum of physically distinct configurations: singular
black holes surrounded by dark-matter halos, regular black holes with
non-singular cores, naked singularities, and Einstein-cluster
solutions. This classification allows us to isolate the role played by
horizon structure, curvature singularities, and core regularization in
governing the onset, persistence, and termination of chaotic motion.

A unified diagnostic framework is employed throughout, based on orbital
projections and Poincar\'e sections applied consistently across all
spacetimes. By maintaining the same dynamical setup and perturbative
structure, we ensure that differences in chaotic behavior can be attributed
directly to geometric properties rather than to modeling artifacts. In
particular, special attention is devoted to the finite-time nature of chaos
in geometries admitting capture, in contrast with the more extended chaotic
phases that may arise in regular configurations.

This section therefore serves a dual purpose. First, it establishes chaos as
a structurally robust feature of EMRI dynamics in non-vacuum environments.
Second, it clarifies how spacetime geometry and dark-matter parameters
modulate its manifestation, thereby providing the physical foundation for
the gravitational-wave analysis presented in the next section.

\subsection{Singular black holes in dark matter halos}
\label{subsec:singular}

We begin with singular black-hole geometries immersed in dark-matter
distributions. These configurations represent the closest extensions of the
Schwarzschild solution and thus provide a natural reference point for
assessing the impact of environmental perturbations on orbital stability.
By first analyzing spacetimes that retain a central curvature singularity,
we establish a baseline against which the effects of core regularization
and horizon modification can later be contrasted.

\subsubsection{Zhao-type dark-matter-embedded black holes}
\label{subsubsec:zhao}

As a first example, we consider a black hole immersed in a dark-matter halo 
described by the Zhao density profile \cite{Zhao:1995cp}
\begin{equation}
\label{eq:zhao_profile}
\rho(r)=\rho_s\,\frac{(r/r_s)^{\mu-3}}{\left[1+(r/r_s)^\nu\right]^{
	(\mu+\alpha)/\nu}} \, ,
\end{equation}
where $r_s$ is a characteristic length scale, $\rho_s$ denotes the central 
density, and $(\alpha,\mu,\nu)$ are dimensionless parameters. In what follows 
we 
restrict ourselves to the representative choice $(\alpha,\mu,\nu)=(1,3,2)$, 
which leads to a well-behaved density profile at large radii.

The corresponding spacetime geometry is assumed to take the form
\begin{equation}
\label{eq:zhao_metric}
\dd s^2 = -c^2 f(r)\,\dd t^2
+ \frac{\dd r^2}{1-\dfrac{2Gm(r)}{c^2 r}}
+ r^2 \dd\Omega^2 \, ,
\end{equation}
with
\begin{equation}
f(r)=1-\frac{2Gm(r)}{c^2 r} \, ,
\end{equation}
where $m(r)$ is the enclosed mass function determined by the dark-matter 
distribution. In Appendix  \ref{secaa} we provide an outlined derivation of the 
metric 
function $f(r)$, which takes the form
\begin{align}
\label{metric1}&m(r)=M - \frac{2 \pi  r_s^4 \rho _s r}{r^2+r_s^2}+2 \pi  r_s^3 
\rho _s \arctan \Big(\frac{r}{r_s}\Big)\,,\\
&f(r) = 1-\frac{2 GM}{c^2r}+\frac{4 \pi G r_s^4 \rho 
_s}{c^2(r^2+r_s^2)}-\frac{4 
\pi G r_s^3 \rho _s}{c^2r} \arctan \Big(\frac{r}{r_s}\Big)\,,	 
\end{align}
where the constant of integration $M$ is the mass of the central object (which, 
as we shall see, is a BH).

Although the Ricci scalar remains finite throughout the spacetime, the 
Kretschmann invariant diverges at the origin, indicating the presence of a 
curvature singularity. Furthermore, the equation $f(r)=0$ admits positive real 
roots, corresponding to the existence of an event horizon. In the limit $\rho_s 
\to 0$, the solution continuously reduces to the Schwarzschild black hole, 
confirming the singular black-hole character of the spacetime.

Concerning the distribution of dark matter inside the horizon, we note that 
there is currently no observational evidence excluding its presence in this 
region. One may either assume that the dark-matter halo extends smoothly from 
spatial infinity down to the singularity, or alternatively introduce a matching 
to an interior vacuum solution at a finite radius. Since our focus lies on the 
dynamics of massive particles in the vicinity of the event horizon, we adopt 
the 
former assumption for simplicity.

\begin{figure}[!htb]
\centering
\vspace{0.1cm}
\includegraphics[width=0.7\linewidth]{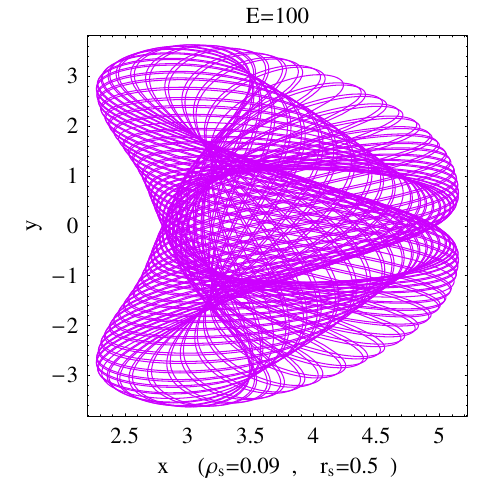}\\
\includegraphics[width=0.7\linewidth]{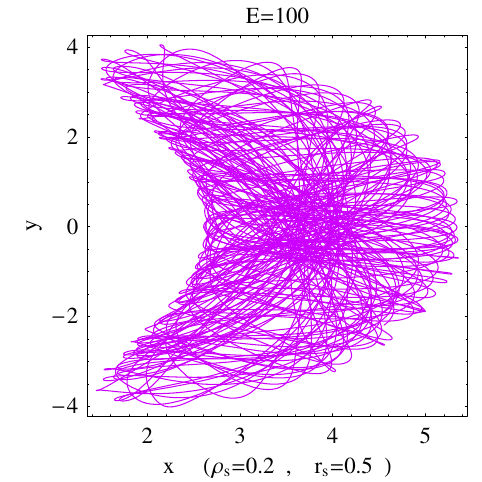}
\caption{{\it{Orbital projections onto the equatorial plane ($\theta=\pi/2$) in the spacetime described by metric~\eqref{metric1}, for $E=100$ and $r_s=0.5$. These are parametric plots of ($r(t)\cos \phi(t),\,r(t)\sin \phi(t)$). The integration interval is $0 \le t \le 6000$ (in units of $M$). Upper panel: Non-chaotic motion for $\rho_s=0.09$, corresponding to horizon radius $r_h=2.16$. Lower panel: Chaotic motion for $\rho_s=0.2$, corresponding to $r_h=2.36$. Units $G=c=M=1$ are adopted. }}}
\label{Figxy100}
\end{figure}

\begin{figure}[!htb]
\centering
\vspace{0.1cm}
\includegraphics[width=0.9\linewidth]{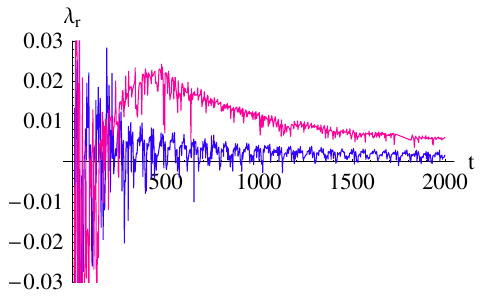}\\
\includegraphics[width=0.9\linewidth]{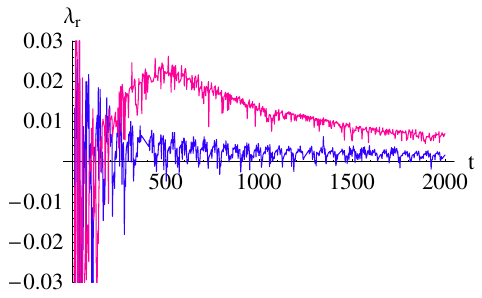}
\caption{{\it{Plots of Lyapunov radial exponent $\lambda_r$ versus $t$ for
		the spacetime described by metric~\eqref{metric1} with $E=100$ $r_s=0.5$. The blue curves correspond to the upper panel of Fig.~\ref{Figxy100} (non-chaotic motion with $\rho_s=0.09$). The magenta curves correspond to the lower panel of Fig.~\ref{Figxy100} (chaotic motion with $\rho_s=0.2$). Upper panel: We took $\delta r(0)=0.00001$. Lower panel: We took $\delta r(0)=0.000001$.
		Units $G=c=M=1$ are adopted. }}}
\label{FigLyap}
\end{figure}

Within this background, orbital motion exhibits both regular and chaotic
regimes, depending on the system parameters (see Fig.~\ref{Figxy100}, where
orbital projections onto the equatorial plane are depicted).
To complement the orbital analysis, in Fig.~\ref{FigLyap} we
present the corresponding evolution of the radial Lyapunov exponent
$\lambda_r$ for the spacetime described by metric~\eqref{metric1} with
$E=100$ and $r_s=0.5$ (see Appendix~\ref{secac} for its definition). The blue
curves correspond to the non-chaotic trajectories shown in the upper panel of
Fig.~\ref{Figxy100}, while the magenta curves correspond to the chaotic
trajectories of the lower panel. As expected, the chaotic configurations
exhibit systematically larger values of the Lyapunov exponent and a more
persistent sensitivity to initial conditions, consistent with the transition
from regular to chaotic motion.

As discussed in Section  \ref{sec:chaos}, 
chaotic behavior is identified through irregular orbital projections and 
stochastic Poincar\'e sections, while regular motion is associated with smooth, 
quasi-periodic structures.


\subsubsection{Bronnikov-type singular configurations}
\label{subsubsec:bronnikov}

As a second example of a singular black hole embedded in a dark-matter 
environment, we consider the spacetime introduced in 
\cite{Balart:2014jia,Bronnikov:2017sgg}, 
characterized by the metric function
\begin{equation}
\label{eq:bronnikov_metric}
f(r)=1-\frac{2GM}{c^2r}
+\frac{8\pi G r_s^4 \rho_s}{c^2r\left(r^3+r_s^3\right)^{1/3}} \, .
\end{equation}
This geometry also possesses a curvature singularity at the origin and admits 
at 
least one event horizon for appropriate parameter choices. Note 
that if $M=4\pi r_s^3\rho_s$, the curvature and Kretschmann scalars 
remain finite for all $r$, and the solution becomes regular (compare with 
Eq.~(25) 
of~\cite{Balart:2014jia}). 

In the above formula, $r_s$ denotes the characteristic scale 
radius of the
dark-matter distribution, while $\rho_s$ determines the corresponding density
scale near the compact object. In the asymptotic large-$r$ regime, the lapse
function~(\ref{eq:bronnikov_metric}) behaves as
\[
f(r)\sim 1-\frac{2GM}{c^2r}+\mathcal{O}(r^{-2})\,,
\]
showing that the spacetime is asymptotically flat and approaches the
Schwarzschild geometry at sufficiently large distances. In the near-core region
$r\sim r_s$, the dark-matter contribution becomes effectively constant and can
significantly modify the local geometry relative to the vacuum case.
Furthermore, for $r\gg r_s$ the dark-matter correction decays rapidly as
$\sim r_s^4/r^2$, implying that environmental effects become negligible far
from the compact object. Finally, setting $M=0$ yields a regular matter-filled
spacetime without curvature singularities.

\begin{figure}[!htb]
\centering
\includegraphics[width=0.7\linewidth]{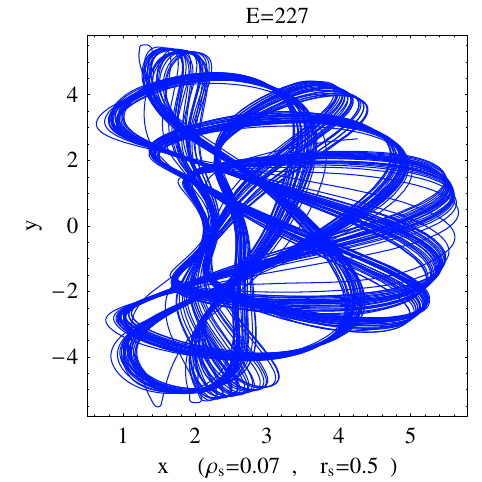}\\
\includegraphics[width=0.7\linewidth]{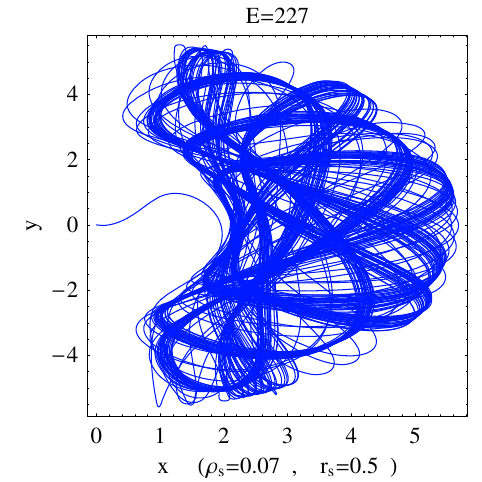}
\caption{{\it{Orbital projections onto the equatorial plane ($\theta=\pi/2$) in 
		the Bronnikov-type geometry~\eqref{eq:bronnikov_metric}, for $E=227$, 
		$\rho_s=0.07$, and $r_s=0.5$. These are parametric plots of ($r(t)\cos \phi(t),\,r(t)\sin \phi(t)$). The corresponding horizon radius is $r_h=1.94$.
		Upper panel: Chaotic trajectory integrated over $0 \le t \le 5000$ (in units of 
		$M$), terminating at $(x,\,y)=(4.26,\, 2.65)$ prior to plunge.
		Lower panel: Chaotic trajectory integrated over $0 \le t \le 5800$, 
		ultimately ending at the central singularity $(x,y)=(0,0)$.
		Units $G=c=M=1$ are used. }}}
\label{Figxy227}
\end{figure}
\begin{figure}[!htb]
\centering
\includegraphics[width=0.73\linewidth]{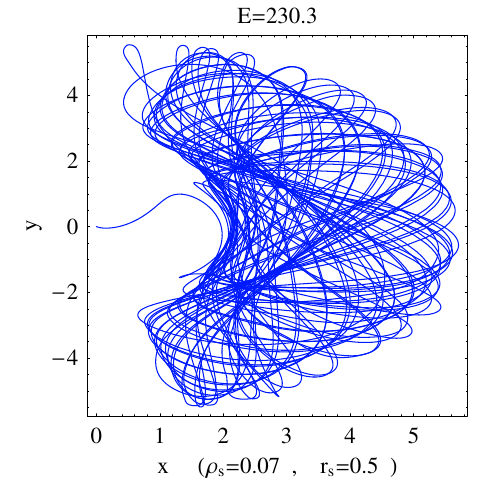}
\caption{{\it{Orbital projection onto the equatorial plane ($\theta=\pi/2$) in 
		the Bronnikov-type spacetime~\eqref{eq:bronnikov_metric}, for $E=230$, 
		$\rho_s=0.07$, and $r_s=0.5$. These are parametric plots of ($r(t)\cos \phi(t),\,r(t)\sin \phi(t)$). The corresponding horizon radius is $r_h=1.94$. 
		The trajectory is integrated over the interval $0 \le t \le 2000$ (in units 
		of $M$). Units $G=c=M=1$ are adopted. }}}
\label{Figxy230p3}
\end{figure}

Compared to the Zhao-type configuration, the Bronnikov-type spacetime exhibits 
a 
richer dynamical structure. For identical values of the external-force 
parameters and of the particle mass, particle trajectories often display a 
transition 
from initially regular motion to chaotic behavior before ultimately terminating 
at the singularity (see Fig. \ref{Figxy227} and Fig. \ref{Figxy230p3}). Such 
mixed dynamics is indicative of non-integrability and 
is consistent with the finite-time chaotic behavior discussed in Section 
\ref{sec:chaos}.

The enhanced complexity of the phase-space structure in this case suggests that 
the detailed form of the dark-matter-induced modification to the metric can 
significantly influence the onset and duration of chaotic motion. Nevertheless, 
the presence of a singular core and an event horizon ensures that chaotic 
behavior remains transient, as trajectories eventually fall into the central 
singularity.

\subsubsection{Qualitative comparison}
\label{subsubsec:singular_comparison}

Both singular black-hole configurations considered above share several common 
features. In particular, the presence of an event horizon and a central 
singularity leads to finite-time chaotic dynamics, with chaos typically 
emerging 
in the strong-field region before capture. At the same time, quantitative 
differences in the transition to chaos and the extent of chaotic regions in 
phase space highlight the sensitivity of the dynamics to the detailed spacetime 
structure induced by the dark-matter distribution.

These singular spacetimes therefore provide a useful reference class for 
assessing the robustness of chaotic behavior, against which regular black holes 
and horizonless configurations will be compared in the following subsections.

\subsection{Regular black holes and the role of core regularization}
\label{subsec:regular}

Regular black hole geometries provide a particularly instructive setting for
probing the impact of core regularization on chaotic dynamics. By removing the
central curvature singularity while preserving the existence of an effective
gravitational potential, such spacetimes allow one to disentangle the effects of
strong-field gravity from those associated with singular behavior. In this
context, regular cores are expected to modify both the onset and the temporal
development of chaos, potentially allowing for extended or smoother transitions
between regular and chaotic motion.

In the following, we focus on a representative regular black hole model and
analyze how core regularization influences the structure of phase space and the
associated orbital dynamics in the presence of environmental perturbations.

\subsubsection{Simpson-Visser regular black holes}
\label{subsubsec:simpson_visser}

\begin{figure*}[!htb]
\centering
\includegraphics[width=0.33\linewidth]{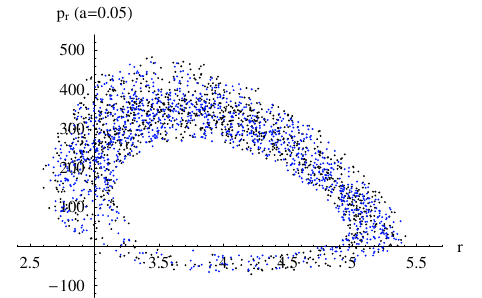}
\includegraphics[width=0.33\linewidth]{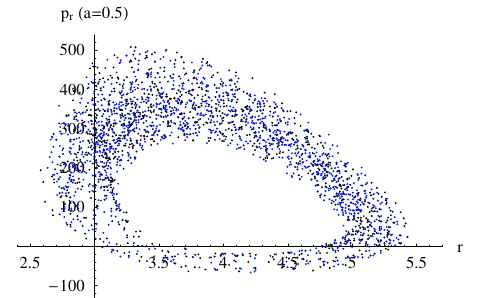}
\includegraphics[width=0.33\linewidth]{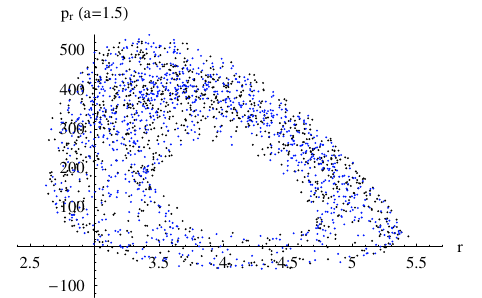}\\
\includegraphics[width=0.33\linewidth]{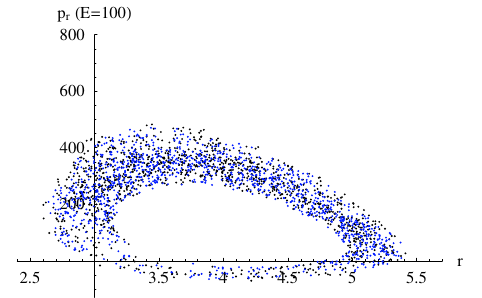}
\includegraphics[width=0.33\linewidth]{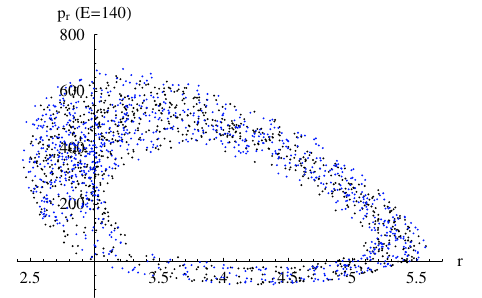}
\includegraphics[width=0.33\linewidth]{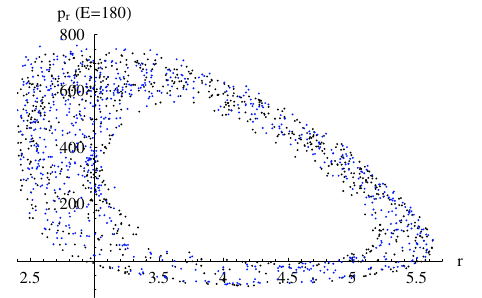}
\caption{{\it{Poincar\'{e} 
		sections constructed at $\phi=0.5$ for the Simpson-Visser 
		spacetime~\eqref{eq:sv_metric}. Initial conditions are 
		$(r(0),\phi(0),p_r(0))=(3.0,0.1,0.1)$ (black points) and $(3.0,-0.1,-0.1)$ 
		(blue points). The integration interval is $0 \le t \le 5\times10^{4}$ (in 
		units of $M$).
		Upper panel: Dependence on the regularization parameter $a$ for fixed $E=100$ 
		and $a=0.05,0.5,1.5$.
		Lower panel: Dependence on the orbital energy $E$ for fixed $a=0.05$ and 
		$E=100,140,180$. Units $G=c=M=1$ are adopted. For relatively small values of the
		regularization parameter $a$ or orbital energy $E$, the trajectories remain
		confined to localized regions of phase space, consistent with predominantly
		regular motion. As $a$ or $E$ increases, the invariant structures become
		progressively distorted and the occupied phase-space region broadens,
		indicating the gradual onset of chaotic dynamics. For sufficiently large
		values of the parameters, the Poincar\'e sections exhibit a substantially more
		irregular and dispersed distribution of points, characteristic of strongly
		non-integrable motion.}}}
\label{Figsvpncr2}
\end{figure*}
As a representative example, we consider the regular black hole proposed by 
Simpson and Visser \cite{Simpson:2018tsi}, described by the line element
\begin{eqnarray}
\label{eq:sv_metric}
&&\dd s^2
=
-c^2\left(1-\frac{2GM}{c^2\sqrt{r^2+a^2}}\right)\dd t^2
+
\frac{\dd r^2}{1-\dfrac{2GM}{c^2\sqrt{r^2+a^2}}}\nonumber
\\
&& \ \ \ \ \ \ \ \ \ \ \     
+
(r^2+a^2)\dd\Omega^2 \, ,
\end{eqnarray}
where $M$ is the mass parameter and $a$ is a length scale that controls the 
regularization of the central region. In the limit $a\to0$, the metric smoothly reduces to the Schwarzschild solution, while for finite $a$ the spacetime is free of curvature singularities. Depending on the value of the parameter $a$, the spacetime may describe either a regular black hole with an event horizon or a horizonless compact object.

For our numerical analysis we have two options, either we work 
with~\eqref{eq:sv_metric} where $g=f$ and
\begin{multline*}
\mathcal{L}
= -mc\sqrt{c^2 f(r) - 2c\sqrt{1-f(r)}\,\dot r - \dot r^{\,2} - (r^2+a^2) 
\dot\phi^{\,2}} \\
- \frac{k_r}{2}\,(r-r_c)^2 - \frac{k_\phi r_h^2}{2}\,\phi^2 \, ,
\end{multline*}
or we bring metric~\eqref{eq:sv_metric} to the 
form~\eqref{eq:sssg}. We choose  the second option and use $\mathcal{L}$ as 
given in~\eqref{eq:lagrangian}. This is done upon introducing the new radial 
variable 
$r_{\text{new}}=\sqrt{r^2+a^2}$. The new functions $g$, $f$, and $h$ read
\begin{align}
&g(r)=\frac{(c^2r-2GM)(r^2-a^2)}{c^2r^3}\,,\qquad r>a\,,\\
&f(r)=\frac{(c^2r-2GM)}{c^2r}\,,\quad h(r)=\sqrt{\frac{f}{g}}=\frac{r}{\sqrt{r^2-a^2}}\,,
%&h(r)=\sqrt{\frac{f}{g}}=\frac{r}{\sqrt{r^2-a^2}}\,,
\end{align}
where we dropped the subscript {\textgravedbl}new{\textacutedbl}.

\subsubsection{Effect of regular core on chaos onset}
\label{subsubsec:regular_core_chaos}

The effect of core regularization on the phase-space structure is displayed in 
Fig.~\ref{Figsvpncr2}. For small values of the regularization parameter $a$, the 
Poincar'e sections remain relatively localized, indicating predominantly regular 
or weakly non-integrable motion. As $a$ increases, the invariant structures 
become progressively distorted and the occupied phase-space region broadens, 
signaling the gradual onset of chaotic behavior. A similar trend is observed 
when the orbital energy $E$ is increased at fixed $a$. Thus, 
Fig.~\ref{Figsvpncr2} shows explicitly that the regular core does not remove 
chaotic dynamics, but modifies the way in which it develops in phase space. This 
behavior contrasts with the singular black-hole configurations discussed 
previously, where chaotic motion is typically terminated by capture at the 
central singularity.

In summary, these findings suggest that regular cores act to moderate the 
abruptness of chaos onset and extend the temporal window over which chaotic 
dynamics can develop. This behavior highlights the importance of distinguishing 
between horizon-related effects and those associated with the central structure 
of the spacetime, a point that will be further clarified in the comparative 
discussion of the following sections.

\begin{figure*}[!htb]
\centering
\includegraphics[width=0.31\linewidth]{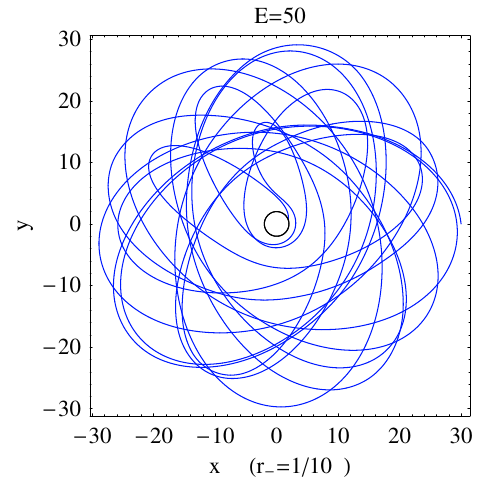}
\includegraphics[width=0.31\linewidth]{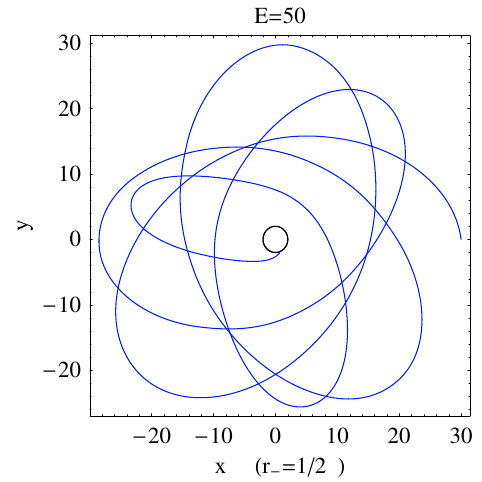}
\includegraphics[width=0.31\linewidth]{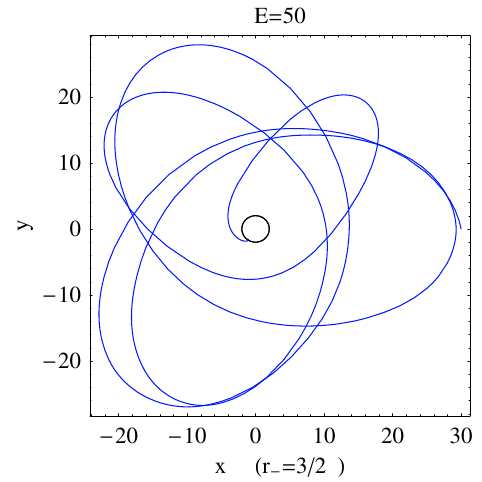}
\caption{{\it{Orbital projections onto the equatorial plane ($\theta=\pi/2$) in 
		the spacetime~\eqref{ns}, corresponding to the case where the central mass $M$ 
		exceeds the dark-matter mass contribution ($M > M_{\rm DM}$). In this regime, 
		the naked singularity is located at $r=r_+$. The trajectories are shown for 
		$E=50$ and $r_- = 0.1,0.5,1.5$. Units $G=c=M=1$ are adopted.}}}
\label{Fignspm2}
\end{figure*}

\begin{figure*}[!htb]
\centering
\includegraphics[width=0.33\linewidth]{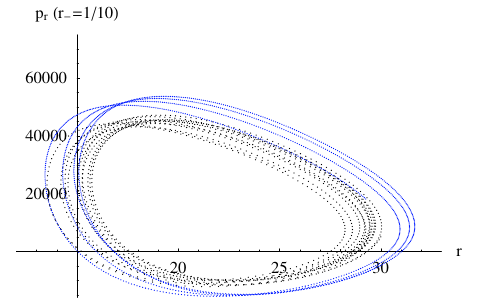}
\includegraphics[width=0.33\linewidth]{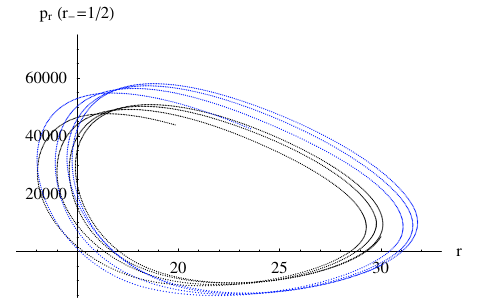}
\includegraphics[width=0.33\linewidth]{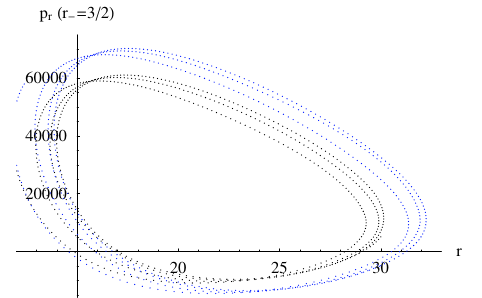}
\caption{{\it{ Poincar\'{e}
		sections in the $(r,p_r)$ plane for the spacetime~\eqref{ns}, corresponding to 
		the regime where the central mass exceeds the dark-matter contribution ($M > 
		M_{\rm DM}$), with the naked singularity located at $r=r_+$. The parameters are 
		$E=50$, $r_+=2GM/c^2=2$, and $r_- = 0.1,0.5,1.5$, illustrating the effect of 
		increasing $r_s^4\rho_s$ for fixed $m$.
		Initial conditions are $(r(0),\phi(0),p_r(0))=(30,0,5000)$ (black points) and 
		$(30,0,-5000)$ (blue points).
		Units $G=c=M=1$ are adopted.
}}}
\label{Fignspncr}
\end{figure*}

\subsection{Naked singularities and Einstein-cluster configurations}
\label{subsec:naked}

We now turn to spacetimes describing compact objects embedded in dark-matter 
distributions that give rise to naked singularities or composite configurations 
involving Einstein clusters. These geometries provide a qualitatively different 
arena for studying chaotic dynamics, as they may lack an event horizon or 
exhibit piecewise structures that interpolate between vacuum and 
matter-dominated regions.

\subsubsection{Metric construction and general properties}
\label{subsubsec:metric4}

We start with configurations that depart qualitatively from the black-hole
geometries considered so far. In particular, we investigate horizonless
spacetimes sourced by an anisotropic dark-matter distribution, which may
describe either naked-singularity configurations or Einstein-cluster
solutions depending on the parameter regime. These geometries provide a
useful laboratory for assessing how the absence of an event horizon modifies
the onset and evolution of chaotic motion.

In Appendix \ref{secab} we outline the derivation of the corresponding metric 
functions
$f(r)$ and $g(r)$ within general relativity.
The spacetime under consideration is characterized by two length scales,
\begin{equation}
r_+ = \frac{2GM}{c^2} \, , \qquad
r_- = \frac{4\pi r_s^4 \rho_s}{M} \, ,
\end{equation}
which encode the contribution of the central mass and the surrounding
dark-matter distribution, respectively. The metric can be written in the
generic spherically symmetric form~(\ref{eq:sssg}), with the explicit
expressions for $f(r)$ and $g(r)$ depending on the relative magnitude of
$r_-$ and $r_+$.

The matter source corresponds to an Einstein cluster, characterized by
vanishing radial pressure and non-zero tangential pressure, while satisfying
the weak energy condition. This provides a physically consistent realization
of an extended dark-matter distribution stretching from a finite radius to
spatial infinity.

\subsubsection{Case $r_- < r_+$: naked singularity configurations}
\label{subsubsec:naked_case1}

We   consider the parameter regime $r_- < r_+$, in which the dark-matter
contribution remains subdominant relative to the central mass scale.
In this case the geometry does not admit an event horizon shielding the
central curvature singularity, and the spacetime represents a naked
singularity immersed in an extended dark-matter distribution.
This configuration provides an instructive contrast with the black-hole
cases discussed previously, since the absence of a horizon can significantly
modify the late-time evolution of chaotic trajectories.

The corresponding metric functions take the form
\begin{align}\label{ns}
&r_+=\frac{2GM}{c^2}\,,\qquad
r_-=\frac{4\pi r_s^4\rho_s}{M}\,,\qquad r_-<r_+\nonumber\\
&m(r)=M\Big(1+\frac{r_-}{r_+}\Big)-M\,\frac{r_-}{r}\,,\nonumber\\
&f(r)=\Big(1-\frac{r_+}{r}\Big)^{\frac{r_+}{r_+-r_-}}
\Big(1-\frac{r_-}{r}\Big)^{-\frac{r_-}{r_+-r_-}}\,,\nonumber\\
&g(r)=\Big(1-\frac{r_+}{r}\Big)
\Big(1-\frac{r_-}{r}\Big)
=\frac{(r-r_+)(r-r_-)}{r^2}\,,
\end{align}
where $M$ denotes the central mass and
\[
\frac{M r_-}{r_+}=\int_{r_+}^\infty 4\pi r^2\rho(r)\dd r
\]
is the total mass contribution from the dark-matter distribution
(with $\rho=m'(r)/(4\pi r^2)=r_s^4\rho_s/r^4$).
The mass function $m(r)$ is defined through
$g(r)=1-2Gm(r)/(c^2r)$.

The radial pressure vanishes identically ($\kappa\equiv 8\pi G/c^4$)
\begin{equation}\label{ns1}
\kappa 
p_r=\frac{c^2r-2Gm(r)}{c^2r^2}~\frac{f'(r)}{f(r)}-\frac{2Gm(r)}{c^2r^3}\equiv 
0\,,
\end{equation}
and the tangential pressure is given by
\begin{equation}\label{ns2}
p_t=\frac{c^2Gm(r)m'(r)}{8\pi r^2[c^2r-2Gm(r)]}>0\,.
\end{equation}
Thus, this dark matter configuration represents   an Einstein cluster. Since 
\begin{align}
&\rho(r)=\frac{m'(r)}{4\pi r^2}=\frac{Mr_-}{4\pi r^4}>0\,,\nonumber\\
&\rho(r)+\frac{p_r}{c^2}>0\,,\nonumber\\
&\rho(r)+\frac{p_t}{c^2}>0\,,
\end{align}
we conclude that the Weak Energy Condition (WEC) is satisfied.

Particle dynamics in this background exhibits finite-time chaotic behavior (see 
Fig.~\ref{Fignspm2}). Trajectories may display chaotic signatures over an 
extended period,  characterized by irregular orbital projections and stochastic 
Poincar\'e  sections, before ultimately terminating at the naked singularity 
(see Fig. \ref{Fignspncr}). The absence of an event horizon implies that 
capture occurs directly at the singularity, 
rendering the chaotic phase intrinsically transient.

The duration and intensity of chaotic motion depend sensitively on the 
parameters controlling the dark-matter distribution. In particular, increasing 
the scale radius of the Einstein cluster tends to enhance the complexity of 
phase-space structures, as reflected in the increased population of both inner 
and outer regions of the Poincar\'e sections prior to capture.

\subsubsection{Case $r_- > r_+$: Einstein cluster surrounding a black hole}
\label{subsubsec:naked_case2}

In the opposite regime, $r_- > r_+$, the dark-matter contribution equals or 
exceeds that of the central mass. The resulting spacetime describes a 
Schwarzschild black hole surrounded by an Einstein cluster extending from 
$r=r_-$ to spatial infinity. In this case, the metric function $g(r)$ remains 
strictly positive outside the horizon, while the interior region $0<r<r_-$ is 
described by the Schwarzschild solution.
\begin{align}\label{bh}
&r_+=\frac{2GM}{c^2}\,,\qquad r_-=\frac{4\pi r_s^4\rho_s}{M}\,,\qquad 
r_->r_+\nonumber\\
&m(r)=2M-M\,\frac{r_-}{r}\,,\nonumber\\
&f=\sqrt{g}\exp\Big\{\sqrt{\frac{r_+}{r_-r_+}}\Big[\arctan\Big(\frac{r-r_+}{
\sqrt{r_+(r_-r_+)}}\Big)-\frac{\pi}{2}\Big]\Big\}\,,\nonumber\\
&g(r)=1-\dfrac{2Gm(r)}{c^2r}=\frac{(r-r_+)^2+r_+(r_- -r_+)}{r^2}>0,
\end{align}
where $M$ is the central mass of the central compact object, and the 
contribution to 
the total mass 
from the DM is also $M=\int_{r_-}^\infty 4\pi r^2\rho(r)\dd r$ (with 
$\rho=m'(r)/(4\pi r^2)=r_s^4\rho_s/r^4$).   The scalar invariant $\mathcal{R}$ 
is given by
\begin{equation}
\mathcal{R}	=\frac{r_+ r_- [2 r^2-6r_+ r+3r_-r_+]}{2 r^4 [(r-r_+)^2+r_+(r_- 
-r_+)]}\,,\qquad r_- >r_+\,,
\end{equation}
and the Kretschmann scalar $\mathcal{K}$ by
\begin{equation}
\mathcal{K}	=\frac{\text{polynomial in }r^6}{r^8 [(r-r_+)^2+r_+(r_- 
-r_+)]^2}\,,\qquad r_- >r_+ \,.
\end{equation}
In the empty region $0<r<r_-$, the solution is the Schwarzschild BH. Thus, the 
piecewise solution, ($f_{\text{BH}}(r),\,g_{\text{BH}}(r)$), given by
\begin{equation}\label{bh1}
f_{\text{BH}}(r)=\begin{cases}
f_0\Big(1-\frac{r_+}{r}\Big) & 0<r<r_-\\
f(r) \text{ as given in }\eqref{bh}& r\geq r_-
\end{cases}\,,
\end{equation}
\begin{equation}\label{bh2}
g_{\text{BH}}(r)=\begin{cases}
1-\frac{r_+}{r} & 0<r<r_-\\
g(r) \text{ as given in }\eqref{bh}& r\geq r_-
\end{cases}\,,
\end{equation}
is a Schwarzschild BH immersed in a DM Einstein cluster. The constant $f_0$ is 
such that
\begin{eqnarray}
&& f_0\sqrt{\frac{r_-r_+}{r_-}}=\nonumber
\\
&&
\exp\Big\{\sqrt{\frac{r_+}{r_-r_+}}\Big[\arctan\Big(\sqrt{\frac{r_-r_+}{r_+}}
\Big)-\frac{\pi}{2}\Big]\Big\}\,.
\end{eqnarray}Finally, the radial pressure vanishes identically for all $r$ and 
the 
tangential 
pressure is given by~\eqref{ns2} for $r>r_-$. We conclude that the WEC is 
satisfied.

\begin{figure}[!htb]
\centering
\includegraphics[width=0.74\linewidth]{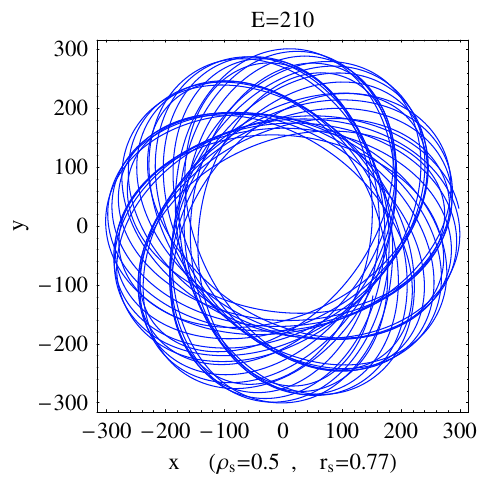}\\
\includegraphics[width=0.74\linewidth]{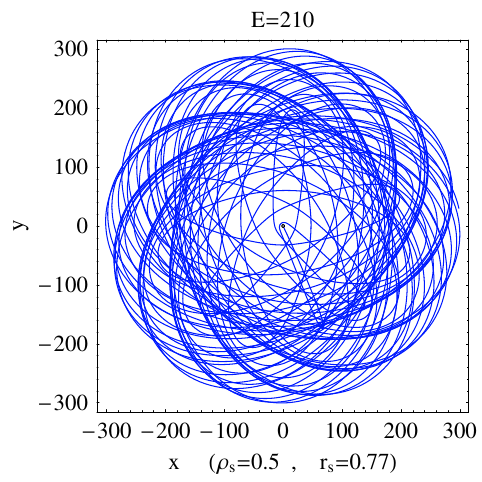}
\caption{{\it{
		Orbital projections onto the equatorial plane ($\theta=\pi/2$) in the spacetime 
		defined by~\eqref{bh1}-\eqref{bh2}, corresponding to the case where the central 
		mass equals the dark-matter mass contribution ($M = M_{\rm DM}$). The 
		parameters are $E=210$, $\rho_s=0.5$, $r_s=0.77$, $r_+=2$, $r_-=2.20873$, and 
		initial radius $r(0)=300$.
		Upper panel: Integration interval $0 \le t \le 5\times10^{4}$.
		Lower panel: Integration interval $0 \le t \le 7\times10^{4}$.
		Units $G=c=M=1$ are adopted. }}}
\label{Figxyif210}
\end{figure}

This piecewise structure leads to a qualitatively different dynamical behavior. 
Particle trajectories typically begin in a regular or quasi-regular regime at 
large radii and subsequently transition to chaotic motion as they approach the 
strong-field region. Chaotic behavior again manifests itself as a finite-time 
phenomenon, with trajectories ultimately crossing the event horizon after 
performing a variable number of loops around the black hole (see 
Fig.~\ref{Figxyif210}).

Notably, variations in the parameters of the Einstein cluster primarily affect 
the duration of the chaotic phase and the number of orbital revolutions 
preceding capture, while the overall qualitative features of the motion remain 
robust. This indicates that, in the presence of an event horizon, the global 
structure of the spacetime plays a dominant role in determining the fate of 
chaotic trajectories.

\subsubsection{Physical interpretation}
\label{subsubsec:naked_interpretation}

The configurations examined in this subsection reveal the interplay between 
horizon structure, matter distribution, and chaotic dynamics. In naked 
singularity spacetimes, the absence of a horizon allows chaotic motion to 
develop until direct capture by the singularity, whereas in Einstein-cluster 
configurations surrounding a black hole, the horizon enforces termination of 
the 
motion after a finite chaotic phase.

These results reinforce the picture that chaotic dynamics in 
dark-matter-embedded spacetimes is generically transient and strongly 
influenced 
by the global causal structure. At the same time, the detailed characteristics 
of chaos, such as its onset, duration, and phase-space morphology, remain 
sensitive to the presence or absence of horizons and to the internal matter 
distribution. These observations will be synthesized and compared across all 
geometries in the following subsection.

\begin{table*}[t]
\centering
\caption{Qualitative comparison of chaotic dynamics in dark-matter-embedded 
spacetimes.}
\label{tab:universality}
\begin{tabular}{lcccc}
\hline\hline
Spacetime type & Horizon & Central structure & Chaos duration & Termination 
mechanism \\
\hline
Singular black hole & Yes & Curvature singularity & Finite-time & Horizon 
crossing \\
Regular black hole & Yes / No & Regular core & Extended finite-time & Horizon 
crossing (if present) \\
Naked singularity & No & Curvature singularity & Finite-time & Direct capture \\
Einstein cluster + BH & Yes & Matter-supported core & Finite-time & Horizon 
crossing \\
\hline\hline
\end{tabular}
\end{table*}

\subsection{Universality and metric dependence of chaotic motion}
\label{subsec:universality}

Having analyzed chaotic dynamics across a broad class of
dark-matter-embedded spacetimes, we can now distinguish between
features that are genuinely universal and those that depend on the
specific geometric realization of the central configuration.
This comparative perspective allows us to separate structural
properties of environmentally perturbed EMRIs from
model-dependent effects tied to horizon structure and internal
geometry.

A first robust conclusion is that chaos emerges generically once
integrability is broken in the strong-field regime.
In all configurations examined, i.e. singular black holes, regular black holes,
naked singularities, and Einstein-cluster spacetimes, the combined
effect of strong gravity and environmental perturbations produces
irregular orbital projections and stochastic Poincar\'e sections,
according to the diagnostics introduced in Section \ref{sec:chaos}.
This demonstrates that chaotic behavior is not an artifact of a
particular metric choice, but a structural consequence of
non-integrable EMRI dynamics in non-vacuum environments.

At the same time, the detailed manifestation of chaos depends
sensitively on global and internal geometric properties.
In singular black-hole geometries, the presence of an event horizon
and a central curvature singularity renders chaos intrinsically
finite in duration, as trajectories inevitably terminate in capture
after a transient chaotic phase.
Regular black holes modify this picture by eliminating the central
singularity, allowing chaotic motion to persist for longer
intervals and producing smoother transitions between regular and
chaotic regimes.
In naked-singularity configurations, chaotic motion may develop
until direct capture by the singularity, whereas in
Einstein-cluster configurations surrounding a black hole, the
horizon again enforces termination after a finite chaotic episode.
Thus, while the \emph{existence} of chaos is robust, its
\emph{lifetime and dynamical morphology} are controlled by the
underlying spacetime structure.

Dark-matter parameters further modulate the onset and extent of
chaotic behavior.
Changes in the scale radius and density primarily affect the energy
threshold for chaos, the duration of the chaotic phase, and the
degree of phase-space mixing.
However, these variations do not alter the qualitative character of
chaos itself, which remains strongly associated with the
strong-field region and typically exhibits finite-time behavior in
geometries admitting capture.

The principal qualitative trends are summarized in
Table~\ref{tab:universality}, which collects the common features and
geometry-dependent differences identified across all cases.
Overall, our analysis establishes chaotic dynamics in
dark-matter-embedded EMRIs as a structurally robust phenomenon,
while demonstrating that its detailed imprint encodes information
about horizon structure, core regularization, and matter
distribution.
This unified perspective provides the conceptual connection to the
gravitational-wave signatures examined in the next section.

\section{Gravitational-Wave Signatures of Chaotic EMRIs}
\label{Gwswction}

Having established the dynamical properties of chaotic motion across a broad
class of dark-matter-embedded spacetimes, we now proceed to the corresponding
gravitational-wave signatures. Our aim in this section is not to assess
detectability or to provide high-precision waveform modeling, but rather to
identify robust qualitative imprints of chaotic dynamics on the emitted
radiation. In this spirit, we adopt a simplified but physically transparent
waveform construction that allows for a direct comparison between regular and
chaotic orbital regimes.

The focus of the following analysis is therefore on how the loss of orbital
regularity manifests itself in the time-domain structure and qualitative
frequency content of gravitational-wave signals. By correlating features of the
waveforms with the underlying phase-space dynamics discussed in the previous
sections, we establish a clear connection between nonlinear orbital motion and
gravitational-wave phenomenology in environmentally perturbed EMRIs.

\subsection{Numerical Kludge waveform construction}
\label{subsec:nk}

In order to investigate the gravitational-wave signatures associated with 
chaotic dynamics, we use the numerical Kludge (NK) approach to waveform 
generation \cite{Babak:2006uv}. This method provides a computationally 
efficient framework for 
constructing approximate gravitational-wave signals from extreme mass-ratio 
inspirals, while retaining the essential features of the underlying orbital 
motion. The gravitational wave signals contain information about both chaotic 
and non-chaotic orbits due to EMRIs, therefore we are looking for any chaotic 
imprints considering different spacetimes.

Within the NK scheme, the trajectory of the small compact object is computed by 
solving the relativistic equations of motion in the curved background 
spacetime, 
as described in Sections \ref{sec:framework} and \ref{sec:metrics}. The 
resulting worldline is then 
interpreted 
as a source moving in flat spacetime, and the emitted gravitational radiation 
is 
calculated using the standard multipolar expansion of the gravitational field, 
typically at the quadrupole level. This hybrid procedure captures the dominant 
kinematical imprints of the motion on the waveform, while avoiding the 
technical 
complexity of fully relativistic waveform generation.

The numerical Kludge approach has been widely used in the literature as a 
qualitative tool for studying EMRI waveforms, particularly in contexts where 
the 
primary goal is to assess the impact of modifications to the orbital dynamics 
rather than to construct high-accuracy templates 
\cite{Shabbir:2025kqh,Zhao:2024exh, Xamidov:2026kqs}. In the present work, this 
framework is especially well suited for isolating the effects of chaotic 
motion, 
since it allows for a direct comparison between waveforms generated by regular 
and chaotic trajectories under otherwise identical conditions.

It is important to emphasize that the NK waveforms employed here are not 
intended to provide faithful representations of signals detectable by future 
space-based observatories. Effects such as gravitational self-force, radiation 
reaction, and strong-field waveform corrections are neglected. Nevertheless, 
the NK approach is sufficient for identifying qualitative differences 
in waveform morphology and spectral content induced by chaotic dynamics, which 
is the primary focus of this study.

We assume that a test object of mass $m$ orbits a gravitational source
(either a black hole or a naked singularity), while its motion is governed
by the effective non-geodesic dynamics introduced in Section II through the
additional environmental perturbations in the Lagrangian.
Under the assumption $M\gg m$, the 
quadropole formula governing the emission of gravitational waves from the EMRI 
source is given by \cite{Shabbir:2025kqh,Zhao:2024exh, Xamidov:2026kqs}
\begin{equation}
h_{ij}=\frac{4\eta M}{D_L}\left(v_i v_j-\frac{m}{r}n_i n_j\right),
\end{equation}
where $D_L$ is the luminosity distance of the EMRI system, 
$\eta=\frac{Mm}{\left(M+m\right)^2}$ represents the symmetric mass ratio, 
$v_i$ denotes the ith component of the spatial velocity of the test particle, 
and $n_i$ is the normalized unit vector pointing in the radial direction 
corresponding to the motion of the test particle.

The gravitational wave signal after projected onto the detector-adapted 
coordinate system, yields the corresponding plus $h_+$ and cross $h_\times$ 
polarizations, given by \cite{Shabbir:2025kqh}
\begin{align}
h_+ 
&=-\frac{2\eta}{D_L}\frac{M^2}{r}
\left(1+\cos^2\iota\right)\cos\left(2\phi+2\zeta\right),\\
h_\times &=-\frac{4\eta}{D_L}\frac{M^2}{r}\cos 
\iota\sin\left(2\phi+2\zeta\right),
\end{align}
where $\iota$ denotes the inclination angle between the orbital angular 
momentum of the EMRI system and the line of sight, while $\zeta$ represents the 
latitudinal angle. To illustrate the corresponding gravitational waveform of 
different regular and chaotic orbits, we implement them within different 
spacetimes in subsequent sections taking $\iota =\zeta =\pi/4$ and $D_L=200$ 
Mpc ($M=1=c=G$ and $m=10^{-5}$ as in the previous sections).

The waveforms presented in the following subsections are constructed
directly from the numerically integrated orbital trajectories.
Specifically, the equations of motion derived in Section \ref{sec:framework} 
are solved
in coordinate time to obtain $r(t)$ and $\phi(t)$, from which the
Cartesian trajectory is reconstructed in the equatorial plane.
The gravitational radiation is then computed using the standard
quadrupole formula within the numerical Kludge approximation,
assuming a distant observer and neglecting radiation-reaction
backreaction on the orbit.
The $+$ and $\times$ polarization modes are obtained from the
second time derivatives of the mass quadrupole moment, and are
plotted as functions of coordinate time in units where $G=c=1$.
This procedure ensures that differences between waveforms reflect
purely dynamical effects associated with regular or chaotic motion.

\subsection{Waveforms from regular and chaotic orbits}
\label{subsec:waveforms}

\begin{figure}[!htb]
\centering
\includegraphics[width=0.99\linewidth]{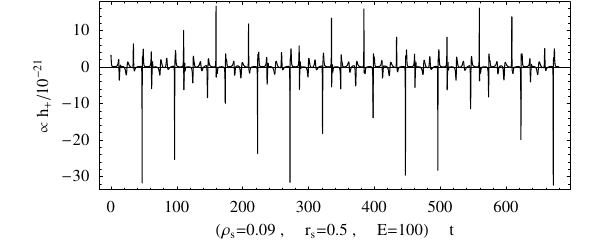}\\
\includegraphics[width=0.99\linewidth]{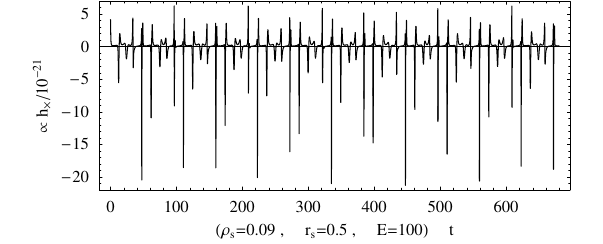}
\caption{{\it{Gravitational-wave signals generated in the regular (non-chaotic) 
		regime for $E=100$, $r_s=0.5$, and $\rho_s=0.09$, corresponding to the upper 
		panel of Fig.~\ref{Figxy100} for metric~\eqref{metric1}. The waveforms exhibit 
		smooth and quasi-periodic amplitude modulation, reflecting the underlying 
		regular orbital motion. Upper panel: $h_+$. Lower panel: $h_\times$. Time $t$ 
		is measured in units of $M$.}}}
\label{Figncpc100}
\end{figure}

\begin{figure}[!htb]
\centering
\includegraphics[width=0.99\linewidth]{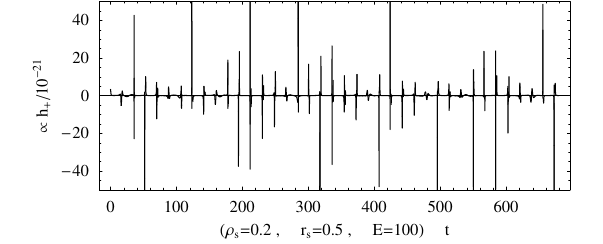}\\
\includegraphics[width=0.99\linewidth]{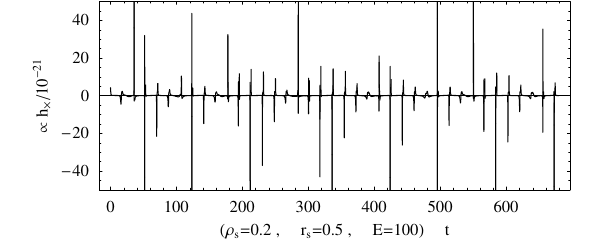}
\caption{{\it{Gravitational-wave signals in the chaotic regime for $E=100$, 
		$r_s=0.5$, and $\rho_s=0.2$, corresponding to the lower panel of 
		Fig.~\ref{Figxy100} for metric~\eqref{metric1}. In contrast to the regular case 
		of Fig. \ref{Figncpc100}, the waveforms display irregular amplitude modulation 
		and loss of quasi-periodic structure, reflecting the stochastic nature of the 
		underlying orbital dynamics. Upper panel: $h_+$. Lower panel: $h_\times$. Time 
		$t$ is measured in units of $M$. }}}
\label{Figcpc100}
\end{figure}

\begin{figure}[!htb]
\centering
\includegraphics[width=1.05\linewidth]{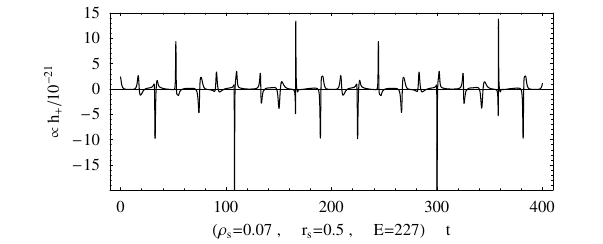}\\
\includegraphics[width=1.05\linewidth]{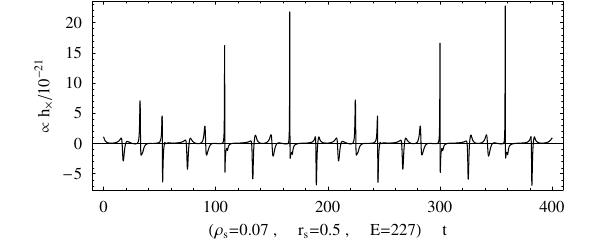}
\caption{{\it{Gravitational-wave signals for $E=227$, $\rho_s=0.07$, and 
		$r_s=0.5$, associated with metric~\eqref{eq:bronnikov_metric} and corresponding 
		to Fig.~\ref{Figxy227}, shown during the early evolution phase $0 \leq t \leq 
		400$ (in units of $M$). In this interval the motion remains nearly regular, and 
		the waveforms exhibit smooth and quasi-periodic amplitude modulation. Upper 
		panel: $h_+$. Lower panel: $h_\times$. }}}
\label{Figcpc227a}
\end{figure}

\begin{figure}[!htb]
\centering
\includegraphics[width=1.05\linewidth]{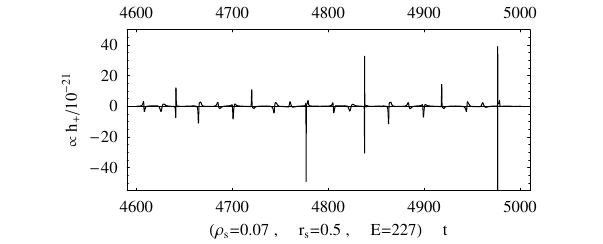}\\
\includegraphics[width=1.05\linewidth]{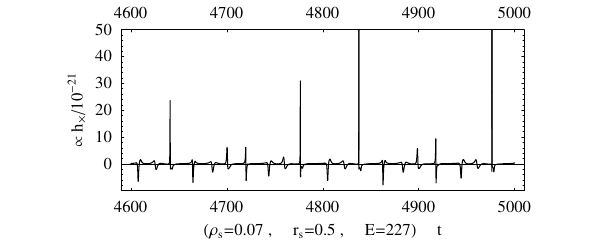}
\caption{{\it{Gravitational-wave signals for $E=227$, $\rho_s=0.07$, and 
		$r_s=0.5$, associated with metric~\eqref{eq:bronnikov_metric} and corresponding 
		to Fig.~\ref{Figxy227}, shown during the late evolution phase $4600 \leq t \leq 
		5000$ (in units of $M$). In this interval the motion has entered a chaotic 
		regime, and the waveforms display irregular amplitude modulation and loss of 
		phase coherence compared to the early-time behavior shown in 
		Fig.~\ref{Figcpc227a}. Upper panel: $h_+$. Lower panel: $h_\times$. }}}
\label{Figcpc227b}
\end{figure}

We now examine the gravitational-wave signals generated by regular and chaotic 
orbital motion, focusing on a direct qualitative comparison between the two 
regimes. In all cases, the waveforms are constructed using the numerical Kludge 
approach described in the previous subsection, and we consider the plus and 
cross polarizations, $h_+(t)$ and $h_\times(t)$, as representative observables 
(see Figs.~\ref{Figncpc100}, \ref{Figcpc100}, \ref{Figcpc227a} 
and~\ref{Figcpc227b}).


For a given background spacetime and fixed external-force parameters, pairs of
trajectories are selected such that one exhibits regular motion while the other
displays chaotic behavior, according to the diagnostics introduced in
Section~\ref{sec:chaos}. All remaining parameters are kept identical, ensuring
that differences in the resulting waveforms can be attributed directly to the
nature of the underlying dynamics.

In the regular regime, the gravitational-wave signals are characterized by a
high degree of coherence. As shown in Fig.~\ref{Figncpc100} for the singular
black-hole spacetime and in Fig.~\ref{Figcpc227a} for the Bronnikov regular
black-hole spacetime during its early evolution phase, both polarizations
exhibit quasi-periodic oscillations with smoothly varying amplitudes and
well-defined phase evolution. The waveform morphology closely reflects the
regularity of the orbital motion, with repeating patterns that persist
throughout the evolution until termination.

In contrast, waveforms generated by chaotic trajectories display markedly
different features. The chaotic waveform shown in Fig.~\ref{Figcpc100}
exhibits irregular amplitude modulation and a less
uniform temporal structure compared to the regular
signal of Fig.~\ref{Figncpc100}. The loss of orbital
coherence manifests itself as a distortion of the otherwise regular
oscillatory pattern, with fluctuations that vary on multiple timescales.
These effects become increasingly pronounced as the chaotic behavior
intensifies.

In configurations where the orbital dynamics transitions from an initially
regular phase to a chaotic one, the corresponding gravitational-wave signal
exhibits a clear temporal change in character. This behavior is illustrated by
Figs.~\ref{Figcpc227a} and~\ref{Figcpc227b}, which correspond to the same
trajectory at different stages of its evolution. During the early,
quasi-regular stage shown in Fig.~\ref{Figcpc227a}, the waveform resembles
that of a purely regular orbit, while in the later stage shown in
Fig.~\ref{Figcpc227b} it acquires the irregular features associated with
chaotic motion. This temporal correspondence provides a direct mapping
between the dynamical evolution of the system and the morphology of the
emitted gravitational radiation.

These qualitative differences between regular and chaotic waveforms are
observed across all spacetime geometries considered in this work, including
singular black holes, regular black holes, and naked-singularity or
Einstein-cluster configurations. While the detailed waveform structure
depends on the specific background and parameter choices, the contrast
between coherent, quasi-periodic signals and irregular, modulated ones,
illustrated by Figs.~\ref{Figncpc100}-\ref{Figcpc227b}, emerges as a robust
signature of chaotic dynamics.

\subsection{Observable imprints of chaos}
\label{subsec:imprints}

We now combine the main observable features imprinted by chaotic dynamics on 
the gravitational-wave signals of extreme mass-ratio inspirals. Although the 
waveforms discussed in the previous subsection are constructed using 
approximate 
methods, they nonetheless allow for the identification of robust qualitative 
signatures associated with the loss of orbital regularity.

A primary manifestation of chaotic motion in the gravitational-wave signal is 
the appearance of irregular amplitude modulation. While regular or 
quasi-regular 
orbits generate waveforms with smoothly varying and predictable amplitudes, 
chaotic trajectories induce fluctuations that vary erratically in time and lack 
a single dominant modulation scale. These amplitude variations reflect the 
sensitive dependence of the orbital motion on initial conditions and the 
resulting irregular evolution of the source’s quadrupole moment.

An important aspect of the systems considered here is that chaotic behavior is 
generically finite in duration. In spacetimes containing an event horizon or a 
naked singularity, chaotic motion typically develops in the strong-field region 
but eventually terminates when the orbiting object is captured. As a result, 
the 
associated gravitational-wave signatures of chaos are transient, appearing only 
during a limited segment of the inspiral. This finite-time nature of chaos 
distinguishes these systems from idealized models exhibiting sustained chaotic 
behavior and has direct implications for observational prospects.

From a detectability standpoint, the transient and parameter-dependent 
character 
of chaotic imprints suggests that their identification in realistic 
gravitational-wave data will be challenging. Nevertheless, the qualitative 
features identified here, i.e. irregular amplitude modulation and increased 
waveform complexity, may serve as indicators of non-integrable dynamics in 
EMRIs embedded 
in non-vacuum environments. Future studies incorporating more accurate waveform 
modeling and data-analysis techniques will be required to assess whether such 
signatures can be isolated in practice.

Overall, while the present analysis does not aim to establish chaos as an 
observable diagnostic in gravitational-wave astronomy, it demonstrates that 
chaotic dynamics leave systematic and interpretable imprints on the emitted 
radiation. These findings provide a conceptual link between nonlinear orbital 
dynamics and gravitational-wave phenomenology, and motivate further 
investigation within more realistic modeling frameworks.

In the following   we discuss the imprints of chaos as the massive 
particle $m$ moves in the four geometries described by the 
metrics~\eqref{metric1}, \eqref{eq:bronnikov_metric}, \eqref{eq:sv_metric},  
and~\eqref{ns}-\eqref{bh} under the effect of the effective harmonic 
potentials given in~\eqref{eq:lagrangian}. 
The qualitative features discussed above manifest themselves consistently across
the different spacetime geometries examined in this work, albeit with distinct
emphases depending on the underlying gravitational and environmental structure.

In dark-matter-embedded singular black-hole configurations of the Zhao type,
regular and chaotic orbital regimes give rise to clearly distinguishable
gravitational-wave morphologies. Regular motion produces coherent $+$ and
$\times$ polarizations with smoothly varying amplitudes, whereas chaotic motion
induces irregular amplitude modulation and loss of phase coherence. The contrast
between these regimes is particularly transparent when background parameters are
held fixed and only the dynamical character of the orbit is altered.

For Bronnikov-type singular configurations, the gravitational-wave signal
reflects the transitional nature of the orbital dynamics. During early stages,
when the motion remains quasi-regular, the waveform closely resembles that of a
regular inspiral. As the orbit enters a chaotic phase in the strong-field 
region,
the signal develops irregular amplitude variations and enhanced complexity,
before terminating when the particle is captured by the singularity. This
temporal correspondence directly maps the onset of chaos in phase space to
distinct changes in waveform morphology.

Regular black holes of the Simpson-Visser type exhibit a markedly different
behavior. Owing to the absence of a central curvature singularity, chaotic 
motion
develops more gradually and typically over much longer integration times. As a
result, the associated gravitational-wave signatures remain largely coherent for
extended periods, with only mild indications of irregularity at lower energies.
Increasing the orbital energy enhances the degree of non-integrability and leads
to progressively more pronounced deviations from regular waveform patterns,
highlighting the stabilizing role of core regularization.

In naked-singularity and Einstein-cluster configurations, chaotic motion is
again intrinsically finite in duration. The gravitational-wave signal displays
irregular modulation during the chaotic phase, followed by abrupt termination as
the particle is either captured by the naked singularity or crosses the event
horizon in the black-hole case. Variations in the dark-matter scale primarily
affect the duration and intensity of the chaotic phase, while the qualitative
structure of the waveform imprints remains robust.

Taken together, these results demonstrate that while the detailed appearance of
gravitational-wave signals depends on the specific spacetime geometry, the
qualitative imprints of chaotic dynamics, such as irregular amplitude 
modulation,
increased waveform complexity, and transient behavior, are generic. The precise
manifestation of these features encodes information about the presence of
horizons, core regularization, and environmental matter distributions, thereby
providing a direct conceptual link between nonlinear orbital dynamics and
gravitational-wave phenomenology.

\section{Conclusions\label{secconc}}

In this work we have investigated the emergence and physical implications of 
chaotic dynamics in extreme mass-ratio inspirals evolving in 
dark-matter-embedded spacetimes. Motivated by growing astrophysical evidence 
that supermassive black holes are surrounded by non-negligible matter 
distributions, and by earlier indications that environmental perturbations can 
induce non-integrable orbital dynamics, our objective was twofold: first, to 
assess the robustness of chaotic motion across a broad class of spacetime 
geometries, and second, to determine whether such dynamics leave systematic 
imprints on the emitted gravitational radiation, potentially 
providing qualitative information on the role of environmental effects and 
dark-matter distributions in the strong-field regime. In the present framework, 
these effects are modeled through effective non-integrable perturbations that 
phenomenologically capture the influence of the surrounding environment on the 
orbital dynamics.

To this end, we developed a unified dynamical framework for test-particle 
motion in spherically symmetric spacetimes, employing horizon-regular 
coordinates that enable a consistent treatment of singular black holes, 
regular compact objects, naked singularities, and Einstein-cluster 
configurations within a single formalism. Environmental effects were modeled 
through effective harmonic perturbations, providing a controlled mechanism for 
breaking integrability while isolating qualitative dynamical features. 
Chaotic behavior was identified using complementary diagnostics, including 
orbital projections and Poincar\'e sections, with particular attention to the 
finite-time character of chaos in strong-field relativistic systems. 
The Poincar\'e sections obtained throughout the analysis
further support this picture by exhibiting the progressive destruction of
regular invariant structures and the emergence of increasingly irregular
phase-space distributions as the system transitions from regular to chaotic
motion.

Our analysis demonstrates that chaotic dynamics is not an artifact of a 
particular metric realization, but rather a generic consequence of 
environmentally perturbed EMRIs once the motion enters the strong-field regime. 
Across all configurations considered, such as singular black holes, regular 
black holes, naked singularities, and Einstein-cluster geometries, chaos 
emerges when 
integrability is sufficiently broken. At the same time, its detailed 
manifestation depends sensitively on the global and internal structure of the 
spacetime. In singular black-hole geometries, the presence of an event horizon 
and a curvature singularity enforces finite-time chaos, with trajectories 
ultimately terminating after a transient chaotic phase. Regular black holes 
modify this picture by removing the central singularity, thereby allowing 
chaotic dynamics to persist over longer intervals and smoothing the transition 
from regular to chaotic motion. Horizonless configurations exhibit chaotic 
behavior until direct capture, while Einstein-cluster geometries again display 
finite-time chaos governed by horizon crossing.

We further examined the gravitational-wave signatures associated with these 
dynamical regimes using the numerical Kludge approach. Although approximate, 
this framework suffices to reveal robust qualitative distinctions between 
waveforms generated by regular and chaotic trajectories. In systems exhibiting a transition from regular to 
chaotic dynamics, the gravitational-wave signal mirrors this evolution, 
developing irregular features precisely when chaos sets in phase space. These 
imprints appear across all spacetime geometries considered, while their 
duration and intensity encode information about horizon structure, core 
regularization, and the dark-matter distribution.

The central conclusion of this work is therefore twofold. First, chaotic 
dynamics in EMRIs embedded in dark-matter environments is a robust and generic 
strong-field phenomenon. Second, although chaos is typically finite in 
duration, it leaves systematic and physically interpretable imprints on the 
gravitational radiation, establishing a conceptual bridge between nonlinear 
orbital dynamics and gravitational-wave phenomenology. While the detectability 
of such signatures in realistic data remains to be quantified, their consistent 
appearance across disparate geometries underscores their physical relevance.

For comparison with the recent works by Das et 
al.~\cite{Das:2025vja,Das:2025eiv}, 
which analyze chaotic dynamics and gravitational-wave signatures in a 
Schwarzschild-like black hole embedded in a Dehnen-type dark-matter halo at 
leading order, our study introduces several extensions. First, we derive and 
employ an exact black-hole solution associated with the more general Zhao 
density profile, of which the Dehnen profile constitutes a special case, thus 
allowing a consistent treatment of the full metric structure without reliance 
on leading-order approximations. Second, whereas 
Refs.~\cite{Das:2025vja,Das:2025eiv} 
focus exclusively on singular Schwarzschild-like geometries, we perform a 
systematic comparative analysis across singular black holes, regular black 
holes, naked singularities, and Einstein-cluster configurations. This broader 
framework enables us to disentangle universal features of environmentally 
induced chaos from those that depend sensitively on horizon structure, 
singularity properties, and core regularization. In this way, the present work 
generalizes previous findings and provides a unified perspective on chaotic 
dynamics and their gravitational-wave imprints in dark-matter-embedded 
compact-object spacetimes.

Several directions for future investigation naturally follow. Incorporating 
radiation reaction and self-force effects would clarify the interplay between 
dissipation and chaos, and determine whether chaotic behavior is enhanced, 
suppressed, or qualitatively altered during realistic inspirals. More accurate 
waveform modeling, combined with dedicated data-analysis strategies, will be 
necessary to assess the observational prospects for detecting chaotic imprints 
with space-based observatories such as LISA. Extending the present framework to 
rotating spacetimes and more realistic environmental models represents an 
essential step toward establishing the full astrophysical relevance of 
chaotic dynamics in extreme mass-ratio inspirals. These investigations are left 
for future projects.

\begin{acknowledgments}
E.N.S. gratefully acknowledges  the 
contribution of 
the LISA Cosmology Working Group (CosWG), as well as support from the COST 
Actions CA21136 -  Addressing observational tensions in cosmology with 
systematics and fundamental physics (CosmoVerse)  - CA23130, Bridging 
high and low energies in search of quantum gravity (BridgeQG)  and CA21106 -  
COSMIC WISPers in the Dark Universe: Theory, astrophysics and 
experiments (CosmicWISPers). 
\end{acknowledgments}

\appendix
\section{Derivation of metric (\ref{metric1})}
\label{secaa}
\renewcommand{\theequation}{A.\arabic{equation}}
\setcounter{equation}{0}

In this Appendix we outline the derivation of the spacetime metric
(\ref{metric1}) sourced by the Zhao-type dark-matter density profile
introduced in the main text. We work within general relativity and assume
a static, spherically symmetric geometry supported by an anisotropic matter
distribution,
\begin{equation}
T^{\mu}{}_{\nu} = \mathrm{diag}(-c^2\rho,\, p_r,\, p_t,\, p_t)\,,
\end{equation}
where $p_r \neq p_t$ in general, as is natural for halo-like configurations.
The metric is written in terms of a mass function $m(r)$ defined through
$g(r)=1-2Gm(r)/(c^2 r)$, and related to the energy density via
\begin{equation}
\rho(r)=\frac{m'(r)}{4\pi r^2}\,.
\end{equation}

The $(t,t)$ component of the Einstein field equations determines the mass
function. Integrating the corresponding equation for the density profile
considered in the main text yields
\begin{equation}
m(r)=M - \frac{2 \pi r_s^4 \rho_s\, r}{r^2+r_s^2}
+2 \pi r_s^3 \rho_s \arctan\!\Big(\frac{r}{r_s}\Big)\,,
\end{equation}
where $M$ denotes the asymptotic mass of the central object.
Substituting this expression into the metric ansatz leads to the lapse
function
\begin{equation}
f(r)=1-\frac{2GM}{c^2 r}
+\frac{4\pi G r_s^4 \rho_s}{c^2 (r^2+r_s^2)}
-\frac{4\pi G r_s^3 \rho_s}{c^2 r}
\arctan\!\Big(\frac{r}{r_s}\Big)\,.
\end{equation}

The remaining independent components of the field equations determine the
radial and tangential pressures.
From the $(r,r)$ equation we obtain
\begin{equation}
p_r(r)=-c^2\rho(r)
=-\frac{c^2 r_s^4\rho_s}{(r^2+r_s^2)^2}\,,
\end{equation}
while the $(\theta,\theta)$ component yields
\begin{equation}
p_t(r)
=-\frac{G m''(r)}{c^2 r}
=\frac{c^2 r_s^4\rho_s (r^2-r_s^2)}{(r^2+r_s^2)^3}\,.
\end{equation}
The stress tensor is therefore anisotropic but completely determined by
the chosen density profile.

Finally, we verify that the weak energy condition (WEC) is satisfied
throughout the spacetime. The energy density is manifestly positive,
\begin{equation}
\rho(r)=\frac{r_s^4\rho_s}{(r^2+r_s^2)^2}>0\,,
\end{equation}
and one finds
\begin{align}
\rho(r)+\frac{p_r}{c^2}&=0\,,\\
\rho(r)+\frac{p_t}{c^2}
&=\frac{2 r_s^4\rho_s\, r^2}{(r^2+r_s^2)^3}>0\,.
\end{align}
Hence the matter configuration sourcing metric (\ref{metric1})
satisfies the weak energy condition everywhere and is therefore
physically admissible.

\section{Derivation of metrics (\ref{ns}) and (\ref{bh})}
\label{secab}
\renewcommand{\theequation}{B.\arabic{equation}}
\setcounter{equation}{0}

In this Appendix we outline the construction of the metrics (\ref{ns}) and
(\ref{bh}), which describe a dark-matter-supported Einstein-cluster
configuration surrounding a central compact object. We consider the general
static, spherically symmetric line element
\begin{equation}\label{a1}
\dd s^2=-c^2 f(r)\dd t^2 + \frac{\dd r^2}{1-\dfrac{2Gm(r)}{c^2 r}} + r^2 
\dd\Omega^2\,,
\end{equation}
where the mass function $m(r)$ is defined through
$g(r)=1-2Gm(r)/(c^2 r)$, and we assume an anisotropic effective fluid
\begin{equation}
T^\mu_{\ \nu}=\text{diag}\!\left(-c^2\rho,\,p_r,\,p_t,\,p_t\right)\,,
\end{equation}
with $\kappa\equiv 8\pi G/c^4$.

We are interested in a configuration with density profile
$\rho\propto r^{-4}$, which yields an Einstein cluster when the radial pressure
vanishes identically. Accordingly, we prescribe the mass function
\begin{equation}\label{a2}
m(r)=c_1-\frac{4\pi r_s^4\rho_s}{r}\,,\qquad (r_{\min}\leq r<\infty)\,,
\end{equation}
where $r_s$ is a scale parameter, $\rho_s$ sets the density scale, and the
constant $c_1$ will be fixed by the asymptotic mass content (central object plus
dark matter).

The $(t,t)$ field equation yields the energy density
\begin{equation}
\rho(r)=\frac{m'(r)}{4\pi r^2}=\frac{r_s^4\rho_s}{r^4}\,,
\end{equation}
while the $(r,r)$ component gives the radial pressure in terms of $f(r)$ and
$m(r)$ as
\begin{equation}\label{pr}
\kappa p_r=
\frac{c^2 r-2Gm(r)}{c^2 r^2}\,\frac{f'(r)}{f(r)}
-\frac{2Gm(r)}{c^2 r^3}\,.
\end{equation}
An Einstein-cluster configuration corresponds to $p_r\equiv 0$.
Integrating \eqref{pr} under this condition yields
\begin{equation}\label{a3}
f(r)=\bigg(\sqrt{\frac{r-r_+}{r-r_-}}\,\bigg)^{c_2}\,
\frac{\sqrt{(r-r_+)(r-r_-)}}{r}\,,
\end{equation}
where the second factor is $\sqrt{g(r)}$, and $c_2$ is an integration constant
given by
\begin{equation}\label{a4}
c_2=\frac{1}{\sqrt{1-\frac{8\pi c^2 r_s^4\rho_s}{c_1^2 G}}}\,.
\end{equation}
The parameters $r_\pm$ are the (real) roots associated with $g(r)$,
\begin{equation}\label{a5}
r_{\pm}=\frac{c_1G\pm \sqrt{G\!\left(c_1^2G-8\pi c^2 
	r_s^4\rho_s\right)}}{c^2}\,,
\end{equation}
and the physical character of the solution depends on the ordering of these two
scales, as discussed below.

\subsubsection{Case $r_-/r_+<1$}

In this case the dark-matter distribution is taken to extend from $r=r_+$ to
spatial infinity. The total dark-matter mass contribution is
\begin{equation}\label{a6}
\int_{r_+}^\infty 4\pi r^2 \rho(r)\,\dd r
=\int_{r_+}^\infty \frac{4\pi r_s^4\rho_s}{r^2}\,\dd r
=\frac{c^2}{2G}\,r_-\,.
\end{equation}
Denoting by $M$ the central mass parameter, we fix $c_1$ by requiring that the
asymptotic mass equals the sum of the central contribution plus the dark-matter
mass. Solving $c_1=M+c^2 r_-/(2G)$ gives
\begin{equation}\label{a7}
c_1= M\Big(1+\frac{2\pi c^2 r_s^4\rho_s}{G M^2}\Big)\,.
\end{equation}
With this identification, the metric functions $f(r)$ and $g(r)$ reduce to the
expressions presented in \eqref{ns}, and the corresponding spacetime describes a
naked singularity located at $r=r_+$, surrounded by a dark-matter Einstein
cluster.

\subsubsection{Case $r_-/r_+>1$}

When $r_->r_+$, it is natural to consider the dark-matter distribution as
extending from $r=r_-$ outwards. In this case the mass contribution of the
cluster becomes
\begin{equation}\label{a8}
\int_{r_-}^\infty 4\pi r^2 \rho(r)\,\dd r
=\int_{r_-}^\infty \frac{4\pi r_s^4\rho_s}{r^2}\,\dd r
=\frac{4\pi r_s^4\rho_s}{r_-}=M\,,
\end{equation}
which implies $c_1=2M$ and hence $m(r)=2M-Mr_-/r$. Integrating \eqref{pr} again
under $p_r\equiv 0$ yields the metric functions given in \eqref{bh}. In this
case the region $0<r<r_-$ is vacuum Schwarzschild, while for $r\geq r_-$ the
solution describes a Schwarzschild black hole immersed in an Einstein-cluster
dark-matter distribution.

For completeness, we note that if one were to evaluate the dark-matter mass by
integrating from $r_+$ instead of $r_-$ in the regime $r_->r_+$, one would be 
led
to a naked-singularity-type solution that is not physically appropriate in the
intermediate region $r_+<r<r_-$.

\section{Definition of Lyapunov radial exponent}
\label{secac}
\renewcommand{\theequation}{C.\arabic{equation}}
\setcounter{equation}{0}

In order to complement the qualitative diagnostics based on orbital
projections and Poincar\'e sections, we also employ a Lyapunov-type
indicator that quantifies the sensitivity of nearby trajectories to
initial conditions. Since the systems considered in the present work
typically exhibit finite-time chaotic behavior due to capture by the
central compact object, we focus on the time evolution of the radial
separation between neighboring trajectories over finite integration
intervals.

Let $\delta r(0)$ denote an infinitesimally small initial radial
separation between two nearby trajectories at $t=0$, and let
$\delta r(t)$ be the corresponding separation at a later time $t$.
We then define the radial Lyapunov exponent through
\begin{equation}
\lambda_r(t)=
\frac{1}{t}\ln\left|
\frac{\delta r(t)}{\delta r(0)}
\right| .
\label{C1}
\end{equation}
In practice, the quantity $\lambda_r(t)$ is evaluated numerically
during the orbital evolution and is used as a finite-time diagnostic
of chaotic behavior. Positive values of $\lambda_r(t)$ correspond to
an exponential divergence of nearby trajectories and therefore indicate
sensitivity to initial conditions, while values close to zero or
negative correspond to regular or weakly chaotic motion.

In the numerical implementation, the nearby trajectories are 
initialized
with identical initial conditions except for a small radial perturbation
$\delta r(0)$. The subsequent evolution of the separation is then
monitored throughout the integration interval. Since the trajectories
may terminate after finite time due to capture by the compact object,
the quantity (\ref{C1}) should be interpreted as a finite-time Lyapunov 
indicator rather than as an asymptotic Lyapunov exponent in the strict 
mathematical sense.

%\newpage

\end{document}